\documentclass[12pt,draftclsnofoot,onecolumn]{IEEEtran}

\IEEEoverridecommandlockouts
\usepackage{indentfirst,flushend}
\usepackage{amssymb,bm,mathrsfs,times,amscd,amsmath,amsthm,amsfonts,bbm}
\usepackage{latexsym}
\usepackage{dsfont,diagbox}
\usepackage{graphicx}
\usepackage{subfigure}
\usepackage{color}
\usepackage{cases}

\usepackage{multirow}
\usepackage[sort]{cite}
\usepackage{multicol}
\usepackage[nooneline,flushleft]{caption2}
\usepackage{clrscode}
\usepackage{algorithm}
\usepackage{psfrag,stfloats,nicefrac}
\usepackage{algorithmic}
\usepackage{supertabular}
\usepackage{makecell}
\allowdisplaybreaks[4]

\begin{document}

\title{Hybrid Precoder Design for Cache-enabled Millimeter Wave Radio Access Networks}
\author{Shiwen He,~\IEEEmembership{Member,~IEEE}, Yongpeng~Wu,~\IEEEmembership{Senior Member,~IEEE}, \\Ju Ren,~\IEEEmembership{Member,~IEEE}, Yongming~Huang,~\IEEEmembership{Senior Member,~IEEE}, \\Robert~Schober,~\IEEEmembership{Fellow,~IEEE}, and Yaoxue~Zhang,~\IEEEmembership{Senior Member,~IEEE}
\thanks{S. He is with the School of Computer, Central South University, Changsha 410083, China. He is also with the School of information Technology, Jiangxi University Of Finance and Economics, West Yuping Road, Nanchang 330032, China. (email: shiwen.he.hn@csu.edu.cn). }
\thanks{Y. Wu is with the Department of Electronic Engineering, Shanghai Jiao Tong University, Minhang 200240, China (email: yongpeng.wu@sjtu.edu.cn).}
\thanks{J. Ren is with the School of Computer, Central South University, Changsha 410083, China. (email: renju@csu.edu.cn).}
\thanks{Y. Huang is with the National Mobile Communications Research Laboratory, School of Information Science and Engineering, Southeast University, Nanjing 210096, China. (email: huangym@seu.edu.cn). }
\thanks{R. Schober is with the Institute for Digital Communication, Friedrich-Alexander University of Erlangen-Nuremberg, Erlangen 91058, Germany (email: robert.schober@fau.de).}
\thanks{Y. Zhang is with the School of Computer, Central South University, Changsha 410083, China. (email: zyx@csu.edu.cn).}
}

\maketitle
\vspace{-.6 in}

\begin{abstract}
In this paper, we study the design of  a hybrid precoder, consisting of an analog and a digital precoder, for the delivery phase of downlink cache-enabled millimeter wave (mmWave) radio access networks (CeMm-RANs). In CeMm-RANs, enhanced remote radio heads (eRRHs), which are equipped with local cache and baseband signal processing capabilities in addition to the basic functionalities of conventional RRHs, are connected to the baseband processing unit via fronthaul links. Two different fronthaul information transfer strategies are considered, namely, hard fronthaul information transfer, where hard information of uncached requested files is transmitted via the fronthaul links to a subset of eRRHs, and soft fronthaul information transfer, where the fronthaul links are used to transmit quantized baseband signals of uncached requested files. The hybrid precoder is optimized for maximization of the minimum user rate under a fronthaul capacity constraint, an eRRH transmit power constraint, and a constant-modulus constraint on the analog precoder. The resulting optimization problem is non-convex, and hence the global optimal solution is difficult to obtain. Therefore, convex approximation methods are employed to tackle the non-convexity of the achievable user rate, the fronthaul capacity constraint, and the constant modulus constraint on the analog precoder. Then, an effective algorithm with provable convergence is developed to solve the approximated optimization problem. Simulation results are provided to evaluate the performance of the proposed algorithms, where fully digital precoding is used as benchmark. The results reveal that except for the case of a large fronthaul link capacity, soft fronthaul information transfer is preferable for CeMm-RANs. Furthermore, surprisingly, hybrid precoding outperforms fully digital precoding with soft fronthaul information transfer for medium-to-large file sizes and fronthaul capacity limited mmWave cloud RANs.
\end{abstract}


\section*{\sc \uppercase\expandafter{\romannumeral1}. Introduction}

The fifth generation (5G) wireless communication network is expected to connect a large number of smart electronic equipments (e.g., smartphones, wearable devices, laptops, machine-to-machine communication devices)~\cite{GlobeMog2014,CMagWang2015}. As a result, future mobile communication systems need to meet more stringent requirements  compared to current systems including higher data rates, higher mobile traffic quality, lower latency, and higher spectrum/energy efficiency~\cite{JSACXiao2017,CMagTran2017}.

Cloud radio access networks (C-RANs) are considered to be a promising architecture for 5G wireless systems to significantly enhance network performance to meet the aforementioned requirements~\cite{AccessGupa2015,SurChec2015,NetwSim2016}. In C-RANs,  most baseband signal processing is performed at the baseband processing unit (BBU) pool, which has a high computation capacity, while the less powerful remote radio heads (RRHs), which are equipped with radio frequency (RF) modules, only perform transmission/reception and compression of radio signals. The RRHs are connected to the BBU via fronthaul links~\cite{TSPPark2013,TSPZhou2016,TVTKang2016,JSACDai2016,SPLPark2017,TWCPan2017,TSPLu2017}. To take the limitation of the fronthaul links into account, joint fronthaul compression and precoding designs were investigated for downlink C-RANs in~\cite{TSPPark2013,TSPZhou2016,TVTKang2016}. The potential of C-RANs to improve the  energy efficiency of wireless networks was investigated in~\cite{JSACDai2016} where the energy consumed by the base station (BS), the RF transmission, and the fronthaul links was minimized. Similarly, to reduce the energy consumption of C-RANs, joint precoding and RRH selection was studied for user-centric green multiple-input multiple-output (MIMO) C-RANs in~\cite{SPLPark2017,TWCPan2017,TSPLu2017}. More recently, the authors of~\cite{TITWang2018} investigated the achievable rate region of downlink C-RANs with RRH cooperation. However, the aforementioned works do not consider the possibility of caching at the RRHs to reduce transmission delay and improve network performance.

To alleviate the fronthaul capacity requirement and reduce the user-perceived latency in C-RANs, recently, an evolved network architecture, referred as fog RAN (F-RAN), was proposed where the RRHs have the ability to store and process signals. These RRHs are referred to as enhanced RRHs (eRRHs)~\cite{ConStoj2014,NetvPeng2016,SysLiu2017}. eRRHs can pre-fetch the most frequently requested files during off-peak traffic periods and store them in their local caches, so that the fronthaul overhead during peak traffic periods is reduced. In this way, lower latency and higher spectral efficiency can be achieved~\cite{CMagWang2014}. The fronthaul-aware design of the eRRH cache placement strategy was studied in~\cite{GlobalPeng2015} with the objective to minimize the average download delay of user requests given the limited eRRH cache capacity. The authors of~\cite{TWCKwak2018} investigated how to cache the files at different caching units with the objective to maximize the average requested content data rates subject to a finite service latency. In~\cite{TWCPark2016}, the joint optimization of cloud and edge precoders was studied for different pre-fetching strategies used for populating the caches of the eRRHs of the F-RAN. The authors of~\cite{TWCTao2016} investigated the joint design of multicast beamforming and eRRH clustering for the delivery phase for fixed pre-fetching strategies, with the goal of minimizing the total transmit power and the fronthaul cost subject to predefined quality-of-service (QoS) requirements. In~\cite{OnlineVu2017}, the authors investigated the joint optimization of user association, data delivery rate, and precoding for MIMO F-RANs. More recently, the authors of~\cite{OnlineDai2018} investigated the BS cache allocation problem for C-RANs employing wireless fronthaul links without taking into account the radio access network.

Although C-RANs and F-RANs can achieve higher network performance by coordinating and centralizing computational tasks at the cloud center, achieving the gigabit-per-second data rates required by 5G networks is still challenging due to the spectrum shortage in the sub-6 GHz frequency bands. On the other hand, in the last few years, millimeter wave (mmWave) communication systems have emerged as a promising candidate for providing order of magnitude improvement in the achievable data rate by exploiting the multi-GHz bandwidth in the range of $30-300$ GHz. Hence, mmWave communication is regarded as a promising option to significantly increase network capacity~\cite{TAPRap2013,AccessRap2013,CSTKutty2016}. To balance the tradeoff between hardware cost/complexity and system performance of mmWave systems, hybrid precoding schemes combining digital precoding with analog precoding have been extensively investigated~\cite{JSTSPHeath2016,AccessHe2017,JSTSPKim2016,TWCEl2014,TWCPark2017,TSPNi2017}. The works in~\cite{JSTSPHeath2016,AccessHe2017,JSTSPKim2016,TWCEl2014,TWCPark2017,TSPNi2017} consider the design of hybrid precoders for point-to-point and downlink multiuser mmWave communication systems.
In~\cite{JSTSPChiang2018}, a hybrid precoder is designed based on small-size coupling matrices between beam patterns which are obtained via beamforming training. Spatial- and frequency-wideband effects in massive MIMO systems are analyzed from an array signal processing point of view in~\cite{TSPWang2018}. Furthermore, the authors of~\cite{OnlineKim2017} have investigated the design of hybrid precoders for C-RAN systems where the RRHs cannot cache data. More recently, the authors of~\cite{TCOMHe2018} propose a two-level transmission scheme to reduce simultaneously the burden on the fronthaul links and delivery latency for cache-enabled radio access network.

In this paper, we investigate the design of hybrid precoders comprising an analog and a digital precoder for the delivery phase of downlink cache-enabled mmWave radio access networks (CeMm-RANs) for two different fronthaul information transfer strategies. The first strategy employs hard fronthaul information transfer (HFIT), where the hard information of the requested uncached files is sent to the eRRHs via the fronthaul links. The second strategy employs soft fronthaul information transfer (SFIT), where a quantized version of the precoded signals of the requested uncached files is sent to the eRRHs via the fronthaul links. Compared to conventional micro-wave MIMO C-RANs~\cite{TSPPark2013,TSPZhou2016,TVTKang2016,JSACDai2016} and F-RANs~\cite{TWCPark2016,TWCTao2016,OnlineVu2017}, for hybrid precoding in CeMm-RANs, there are two fundamental differences, namely the constant-modulus constraint on the analog precoder, which is implemented using phase shifters, and the cascading of a digital precoder and an analog precoder. These differences make the design of hybrid precoders very challenging and motivate the investigation of the design of such precoders for the delivery phase of downlink CeMm-RANs in this paper. The main contributions of this paper can be summarized as follows:
\begin{itemize}
\item We investigate the design of hybrid precoders for CeMm-RANs employing HFIT and SFIT, respectively. We jointly optimize the analog precoders and the digital precoders to maximize the minimum user rate in the network.
\item Two effective optimization algorithms are developed to solve the formulated optimization problem. For optimization of the digital precoder for a given analog precoder, we convexify the non-convex analytical expressions for the user rate and the fronthaul capacity constraint via successive convex approximation methods. Then, an iterative algorithm, which is proved to converge to a stationary point, is proposed to solve the resulting optimization problem. For optimization of the analog precoder, we use a Taylor series expansion to optimize the phases of the analog precoder\footnote{We note that the considered optimization problem is more difficult than the related problems in~\cite{TWCEl2014,TWCPark2017}. The authors of~\cite{TWCEl2014,TWCPark2017} investigated the maximization of the spectral efficiency of point-to-point links under a single transmit power constraint. This allows the use of Orthogonal Matching Pursuit (OMP)~\cite{TWCEl2014} and matrix factorization~\cite{TWCPark2017} to obtain the analog precoder. However, these methods cannot be directly applied to the problem considered in this paper.}. A corresponding effective iterative algorithm is developed. The convergence and computational complexity of the presented algorithms are analyzed.
\item Numerical experiments are carried out to evaluate the performance of the presented algorithms. Our results unveil that \romannumeral1) when the system performance is constrained by the fronthaul capacity or the file size, for HFIT, hybrid precoding and fully digital precoding achieve almost the same minimum user rate, \romannumeral2) if the fronthaul capacity is large, HFIT with an approriate number of coordinated eRRHs can outperform SFIT, otherwise, SFIT yields a higher performance, and \romannumeral3) for medium-to-large file sizes, hybrid precoding with SFIT outperforms fully digital precoding with SFIT for fronthaul capacity limited mmWave C-RANs\footnote{MmWave C-RANs can be regarded as a special case of CeMm-RANs where the eRRHs do not cache any files at the local cache.}.
\end{itemize}

The remainder of this paper is organized as follows. The system model is described in Section \uppercase\expandafter{\romannumeral2}. In Section \uppercase\expandafter{\romannumeral3} and \uppercase\expandafter{\romannumeral4}, the design of  hybrid precoding schemes for CeMm-RANs with HFIT and SFIT is investigated, respectively. In Section \uppercase\expandafter{\romannumeral5}, the performance of the developed algorithms is evaluated via simulation. Conclusions are drawn in Section \uppercase\expandafter{\romannumeral6}.

\textbf{Notations}: Bold lower case and upper case letters represent column vectors and matrices, respectively. The superscripts $\left(\cdot\right)^{T}$, $\left(\cdot\right)^{*}$, and $\left(\cdot\right)^{H}$ represent the transpose, the conjugate, and the conjugate transpose operators, respectively. $\mathrm{tr}\left(\cdot\right)$, $\|\cdot\|_{2}$, and $\|\cdot\|_{\mathrm{F}}$ denote the trace, the Euclidean norm, and the Frobenius norm, respectively. $\mathrm{diag}\left(\mathbf{a}\right)$ is a diagonal matrix whose diagonal elements are the elements of vector $\mathbf{a}$. $\mathbf{A}\succeq \bm{0}$ is a positive semidefinite matrix. $\mathbf{0}_{N\times N}$ and $\mathbf{I}_{N\times N}$ denote the $N\times N$ zero and identity matrices, respectively. $\left[\mathbf{A}\right]_{\left(m,n\right)}$ represents the element in row $m$ and column $n$ of matrix $\mathbf{A}$ and $\mathrm{vec}\left(\mathbf{A}\right)$ denotes the column vector obtained by stacking the columns of matrix $\mathbf{A}$ on top of one another. $\circ$ and $\circledast$ denote the Hadamard product and the Kronecker product, respectively. The function $\lfloor x\rfloor$ rounds $x$ to the nearest integer not larger than $x$. $\overline{a}$ denotes the complement $1-a$ of a binary variable $a\in\left\{0,1\right\}$. $\mathbb{E}\left[\cdot\right]$ denotes the expectation operator. $\log\left(\cdot\right)$ is the logarithm with base $e$. $j$ is the imaginary unit, i.e., $j^{2}=-1$. $\mathbb{R}$ and $\mathbb{C}$ are the fields of real and complex numbers, respectively. For a set $\mathcal{A}$, $\left|\mathcal{A}\right|$ denotes the cardinality of the set, while for a complex number $x$, $\left|x\right|$ denotes the absolute value of $x$. The circularly symmetric complex Gaussian distribution with mean $\mathbf{u}$ and covariance matrix $\mathbf{R}$ is denoted by $\mathcal{CN}\left(\mathbf{u}, \mathbf{R}\right)$. The symbols used frequently in this paper are summarized in Table~\ref{SymbolSummarized}.
\begin{table*}[htbp]
	\setlength{\abovecaptionskip}{0pt}
	\setlength{\belowcaptionskip}{5pt}
	\captionstyle{flushleft}
	\onelinecaptionstrue
	\centering
	\caption{Symbols used frequently in this paper.}
	\begin{tabular}{|c|l|}
		\hline
		Symbol&\makecell[c]{Meaning}\\
		\hline
		\hline
		$K_{\mathrm{U}}$, $K_{\mathrm{R}}$ &Numbers of users and eRRHs\\
		\hline
		$\mathcal{K}_{\mathrm{U}}$, $\mathcal{K}_{\mathrm{R}}$ &Sets of users and eRRHs\\
		\hline
		$N_{\mathrm{uRF},k}$, $N_{\mathrm{tRF},i}$ &Numbers of RF chains of user $k$ and eRRH $i$\\
		\hline
		$N_{\mathrm{tt}}$, $N_{\mathrm{tRF}}$& Total numbers of antennas and RF chains at eRRHs\\
		\hline
		$N_{\mathrm{r},k}$, $N_{\mathrm{t},i}$ &Numbers of antennas of user $k$ and eRRH $i$\\
		\hline
		$\mathcal{N}_{\mathrm{R},i}$, $\mathcal{N}_{\mathrm{C},i}$&Sets of antennas and RF chains at eRRH $i$\\
		\hline
		$P_{i}$ &Maximum transmit power of eRRH $i$\\
		\hline
		$C_{i}$ &Capacity of fronthaul link to eRRH $i$\\
		\hline
		$B_{i}$ &Normalized cache size of eRRH $i$\\
		\hline
		$F$ &Number of files in the library\\
		\hline
		$\mathcal{F}$, $\mathcal{F}_{\mathrm{req}}$ &Sets of all files and all requested files\\
		\hline
		$S$, $\overline{S}$ &Normalized size of file and subfile\\
		\hline
		$\mathcal{L}$ &Set of subfiles\\
		\hline
		$L$ &Number of subfiles\\
		\hline
		$f_{k}$ &File requested by user $k$\\
		\hline
		$c_{f,l,i}$ &Binary cache variable of subfile $\left(f,l\right)$ at eRRH $i$\\
		\hline
		$d_{f,l,i}$ &Binary transfer variable of subfile $\left(f,l\right)$ to eRRH $i$\\
		\hline
		$d_{f,l}$ &Number of the spatial data streams of subfile $\left(f,l\right)$\\
		\hline
		$\mathcal{N}_{\mathcal{F},i}$ &Set of subfiles available at eRRH $i$\\
		\hline
		$R_{f,l}$ &Data delivery rate of subfile $\left(f,l\right)$\\
		\hline
		$R_{\mathrm{min}}$ &Minimum user rate\\
		\hline
		$\mathcal{R}$ & $\mathcal{R}=\left\{R_{\mathrm{min}}\right\}\cup\left\{R_{f,l}\right\}_{f\in\mathcal{F}_{\mathrm{req}},l\in\mathcal{L}}$\\
		\hline
		$\mathbf{y}_{k}$ &Signal received by user $k$\\
		\hline
		$\mathbf{x}_{i}$ &Signal transmitted by eRRH $i$\\
		\hline
		$\mathbf{H}_{k,i}$ &Channel matrix from eRRH $i$ to user $k$\\
		\hline
		$\sigma_{k}^{2}$ & Noise variance at user $k$\\
		\hline
		$N_{c}$ &Number of scatterers\\
		\hline
		$\mathbf{F}_{\mathrm{RF},i}$ &Analog precoding matrix at eRRH $i$\\
		\hline
		$\mathbf{G}_{f,l,i}$ &Digital precoding matrix for basedband signal $\mathbf{s}_{f,l}$ at eRRH $i$\\
		\hline
		$\mathcal{G}$ &Set of digital precoding matrices at eRRHs\\
		\hline
		$\mathbf{U}_{f,l,i}$ &Digital precoding matrix for basedband signal $\mathbf{s}_{f,l}$ at BBU for eRRH $i$\\
		\hline
		$\mathcal{U}$ &Set of digital precoding matrices at BBU\\
		\hline
		$\mathbf{\Xi}_{k,l}$ &MMSE receiver for subfile $\left(f_{k}, l\right)$\\
		\hline
		$\mathbf{\Upsilon}_{k,l}$ &Weight matrix for subfile $\left(f_{k}, l\right)$\\
		\hline
		$\mathcal{X}$, $\mathcal{V}$ & Sets of MMSE receivers and weight matrices\\
		\hline
		$\mathbf{\Omega}_{i}$ &Quantization noise covariance matrix to eRRH $i$\\
		\hline
		$\mathcal{O}$ &Set of quantization noise covariance matrices\\
		\hline
	\end{tabular}
	\label{SymbolSummarized}
\end{table*}

\section*{\sc \uppercase\expandafter{\romannumeral2}. System Model}

\begin{figure}[t]
\renewcommand{\captionfont}{\footnotesize}
\renewcommand*\captionlabeldelim{.}
	\centering
	\captionstyle{flushleft}
	\onelinecaptionstrue
	\includegraphics[width=0.8\columnwidth,keepaspectratio]{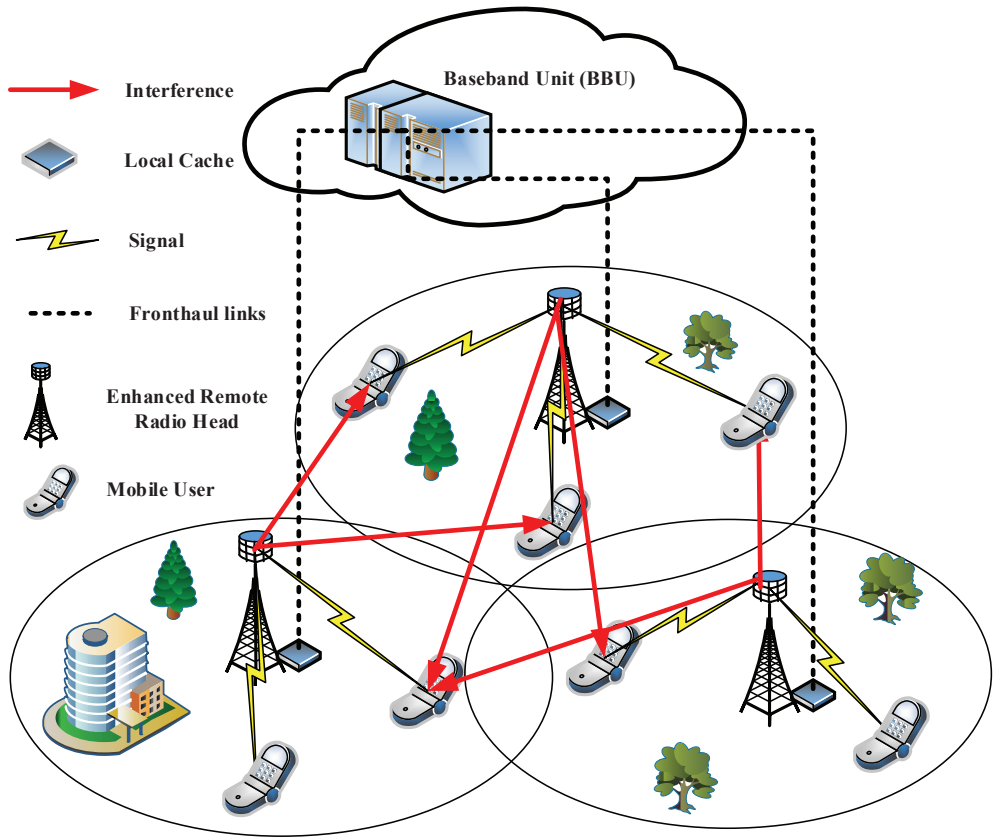}\\
	\caption{Illustration of a CeMm-RAN.}
	\label{EquivalentSystemModel}
\end{figure}

\begin{figure}[ht]
\renewcommand{\captionfont}{\footnotesize}
\renewcommand*\captionlabeldelim{.}
	\centering
	\captionstyle{flushleft}
	\onelinecaptionstrue
	\includegraphics[width=0.8\columnwidth,keepaspectratio]{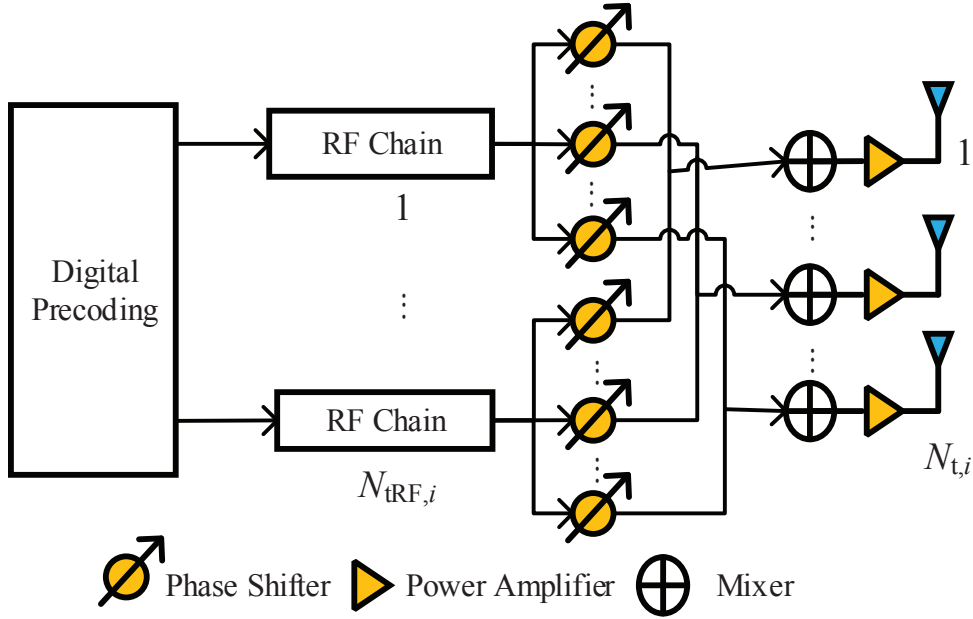}\\
	\caption{Antenna architecture of eRRH $i$, $\forall i\in\mathcal{K}_{\mathrm{R}}$.}
	\label{AntennasArchitecture}
\end{figure}

Consider the downlink of a CeMm-RAN, as illustrated in Fig.~\ref{EquivalentSystemModel}, where $K_{\mathrm{U}}$ multi-antenna users establish wireless communication with $K_{\mathrm{R}}$ eRRHs.
We assume that there are no communication links between the eRRHs. eRRH $i$, $ i\in\mathcal{K}_{\mathrm{R}}=\left\{1,\cdots,K_{\mathrm{R}}\right\}$, is connected to the BBU through an error-free fronthaul link of capacity $C_{i}$ bit/symbol and is equipped with a cache, which can store $nB_{i}>0$ bits, where $n$ is the number of symbols of each downlink coded transmission interval and $B_{i}$ is the normalized cache size~\cite{TWCPark2016}. Furthermore, eRRH $i$ is equipped with $N_{\mathrm{tRF},i}$ RF chains and $N_{\mathrm{t},i}$ transmit antennas. Each RF chain is connected to the $N_{\mathrm{t},i}$ transmit antennas via $N_{\mathrm{t},i}$ phase shifters, $i\in\mathcal{K}_{\mathrm{R}}$, as shown in Fig.~\ref{AntennasArchitecture}.
Each user $k$, $k\in\mathcal{K}_{\mathrm{U}}=\left\{1,\cdots,K_{\mathrm{U}}\right\}$, is equipped with $N_{\mathrm{uRF},k}$ RF chains and $N_{\mathrm{r},k}$ receive antennas. Each RF chain is connected with the $N_{\mathrm{r},k}$ receive antennas via $N_{\mathrm{r},k}$ phase shifters.
\subsection*{A. Channel Model}
The signal received at user $k$ can be expressed as
\begin{equation}\label{Cachenable01}
\mathbf{y}_{k}=\sum\limits_{i\in\mathcal{K}_{\mathrm{R}}}\mathbf{H}_{k,i}\mathbf{x}_{i}+\mathbf{n}_{k},
\end{equation}
where $\mathbf{H}_{k,i}\in\mathbb{C}^{N_{\mathrm{r},k}\times N_{\mathrm{t},i}}$ denotes the channel matrix between user $k$ and eRRH $i$. $\mathbf{x}_{i}\in\mathbb{C}^{N_{\mathrm{t},i}\times 1}$ denotes the signal transmitted by eRRH $i$ and $\mathbf{n}_{k}\sim\mathcal{CN}\left(\mathbf{0},\sigma_{k}^{2}\mathbf{I}_{N_{\mathrm{r},k}\times N_{\mathrm{r},k}}\right)$ is the additive white Gaussian noise (AWGN). Note that in~\eqref{Cachenable01}, we implicitly assume perfect time and frequency synchronization.

Since mmWave channels have limited numbers of scatterers~\cite{TAPRap2013,AccessRap2013,CSTKutty2016}, we adopt a narrowband clustered channel model with $N_{\mathrm{c}}$ scatterers based on the extended Saleh-Valenzuela model. Each scatterer is further assumed to contribute a single propagation path to the channel between a user and an eRRH~\cite{JSTSPHeath2016,JSTSPKim2016,TWCEl2014}. Thus, channel matrix $\mathbf{H}_{k,i}$ can be modeled as
\begin{equation}\label{Cachenable02}
\mathbf{H}_{k,i}=\sqrt{\rho_{k,i}N_{\mathrm{t},i}N_{\mathrm{r},k}}\sum\limits_{p=1}^{N_{\mathrm{c}}}
\alpha_{p,k,i}\mathbf{a}_{\mathrm{u}}\left(\theta_{p,k,i}\right)\mathbf{a}_{\mathrm{r}}^{H}\left(\phi_{p,k,i}\right), \forall k\in\mathcal{K}_{\mathrm{U}}, i\in\mathcal{K}_{\mathrm{R}},
\end{equation}
where $\theta_{p,k,i}\in\left[0,2\pi\right)$ and $\phi_{p,k,i}\in\left[0,2\pi\right)$ are the azimuth angles of departure and arrival (AoDs/AoAs) of the $p$-th path between eRRH $i$ and user $k$, respectively. $\rho_{k,i}$ denotes the average path-loss between eRRH $i$ and user $k$. $\alpha_{p,k,i}$ represents the complex gain of the $p$-th path between eRRH $i$ and user $k$. The path amplitudes are assumed to be Rayleigh distributed, i.e., $\alpha_{p,k,i}\sim\mathcal{CN}\left(0,\sigma_{k,i}^{2}\right)$ with $\sigma_{k,i}^{2}$ being the average power gain, $p\in\mathcal{K}_{\mathrm{C}}=\left\{1,\cdots,N_{\mathrm{c}}\right\}$, $k\in\mathcal{K_{\mathrm{U}}}$, $i\in\mathcal{K_{\mathrm{R}}}$. Assume that a uniform linear array (ULA) is adopted at the eRRHs and the users. In particular, for an $N_{\mathrm{t},i}$-element ULA, the array response vector is given by~\cite{AntennaTheory1997}
\begin{equation}\label{Cachenable03}
\bm{a}_{\mathrm{r}}\left(\phi_{p,k,i}\right)=\sqrt{\frac{1}{N_{\mathrm{t},i}}}
\left[1,e^{j\frac{2\pi}{\lambda_{s}}d_{\mathrm{a}}\sin\left(\phi_{p,k,i}\right)},\cdots,
e^{j\left(N_{\mathrm{t},i}-1\right)\frac{2\pi}{\lambda_{s}}d_{\mathrm{a}}\sin\left(\phi_{p,k,i}\right)}\right]^{T},
\end{equation}
where $\lambda_{s}$ and $d_{\mathrm{a}}$ are the signal wavelength and the antenna spacing, respectively. The array response vector $\mathbf{a}_{\mathrm{u}}\left(\theta_{p,k,i}\right)$ at user $i$ can be written in a similar fashion\footnote{To establish a performance upper bound for hybrid precoding in CeMm-RANs, in this paper, we assume that perfect channel state information (CSI) is available. The CSI can be acquired by the eRRHs and reported back to the BBU via the fronthaul links, e.g.~\cite{TSPZhou2016,TVTKang2016,JSACDai2016,SPLPark2017}, \cite{TWCPark2016,OnlineVu2017,TWCTao2016}. The analysis of the effect of imperfect CSI is an interesting topic for future work.}. 

\subsection*{B. Cache Model}
Assume that each user $k\in\mathcal{K}_{\mathrm{U}}$ requests contents or files from library $\mathcal{F}=\left\{1,\cdots,F\right\}$ stored at the BBU. Without loss of generality, we assume that all files in library $\mathcal{F}$ have a size of $nS$ bits, i.e., $S$ is the normalized file size. In this work, we assume that each file $f\in\mathcal{F}$ is split into $L$ subfiles of equal size, $n\overline{S}=nS/L$ bits\footnote{The proposed algorithms can be applied to the case where the file is not divided into multiple subfiles by setting the value of $L$ to one.}. Based on long-term information regarding the popularity distribution of the files, the cache capacity, the file size, and the fronthaul capacity, each eRRH $i\in\mathcal{K_{\mathrm{R}}}$ pre-stores $nB_{i}$ bits of $nFS$ bits in the library in its cache~\cite{TWCPark2016,OnlineVu2017,TWCTao2016}.  Namely, each eRRH $i\in\mathcal{K}_{\mathrm{R}}$ selects $\lfloor B_{i}/\overline{S} \rfloor$ subfiles from library $\mathcal{F}$ at the BBU and stores them in its local cache.  The cache status of subfile $\left(f,l\right)$, $f\in\mathcal{F}$, $l\in\mathcal{L}=\left\{1,\cdots,L\right\}$, can be modeled by defining binary variables $c_{f,l,i}$, $f\in\mathcal{F}$, $l\in\mathcal{L}$, $i\in\mathcal{K}_{\mathrm{R}}$, as
\begin{equation}\label{Cachenable04}
c_{f,l,i}=
\begin{cases}
	1, ~~\text{if subfile} \left(f,l\right) \text{is cached by eRRH}~i, \\
	0 , ~~\text{otherwise} .
\end{cases}
\end{equation}
For simplicity, in this work, we assume that the cache state information $c_{f,l,i}$, $f\in\mathcal{F}$, $l\in\mathcal{L}$, $i\in\mathcal{K}_{\mathrm{R}}$, is predetermined\footnote{The impact and optimality of the applied caching and content distribution strategies are not studied in this paper. The investigated transmission scheme can be applied for any caching strategy. However, different caching strategies will lead to different system performances~\cite{TWCPark2016,TWCTao2016}, of course.}. Cached files requested by the users can be retrieved directly from the local cache of the serving eRRHs instead of from the BBU. In contrast, uncached files need to be transferred to eRRH $i$ via the fronthaul link. For the information transfer of the uncached files over the fronthaul links to the eRRHs, two different approaches may be distinguished, namely, HFIT and SFIT. In the following, we discuss the design of hybrid precoding for CeMm-RANs with HFIT and SFIT, respectively.

\section*{\sc \uppercase\expandafter{\romannumeral3}. Hybrid Precoding Design for Hard Fronthaul Information Transfer}
In this section, we investigate the design of hybrid precoding for CeMm-RANs with HFIT, where the hard information regarding the subfiles that are not cached at the eRRHs is transferred via the fronthaul links. To facilitate the presentation, in the sequel, we assume that user $k\in\mathcal{K}_{\mathrm{U}}$ independently requests a random file $f_{k}$ from library $\mathcal{F}$. We further assume that only a single file is transmitted to each user in a given transmission interval, as in~\cite{TWCPark2016,TWCTao2016,OnlineVu2017}. Let $\mathcal{F}_{\mathrm{req}}=\bigcup_{k\in\mathcal{K}_{\mathrm{U}}}\left\{f_{k}\right\}$ be the set of files requested by  the $K_{\mathrm{U}}$ users. We define the binary variable $d_{f,l,i}$, $f\in\mathcal{F}$, $l\in\mathcal{L}$, $i\in\mathcal{K}_{\mathrm{R}}$, as%
\begin{equation}\label{Cachenable05}
d_{f,l,i}=
\begin{cases}
1,~~\text{if subfile} \left(f,l\right) \text{is transferred to eRRH}~i,\\
0,~~\text{otherwise} .
\end{cases}
\end{equation}
In this paper, we assume that the values of $d_{f,l,i}$, $f\in\mathcal{F}_{\mathrm{req}}$, $l\in\mathcal{L}$, $i\in\mathcal{K}_{\mathrm{R}}$, are pre-determined. As an example, in the numerical results in Section~{\sc \uppercase\expandafter{\romannumeral5}}, we will assume that variables $d_{f,l,i}$ are set such that uncached subfile $\left(f_{k}, l\right)$ requested by user $k$ is transferred to the $N_{\mathrm{F}}$ eRRHs that have not cached the subfile and have the largest channel gains $\left\|\mathbf{H}_{k,i}\right\|_{\mathrm{F}}^{2}$ to user $k$, where $N_{\mathrm{F}}\leq K_{\mathrm{R}}$ is a tunable parameter~\cite{TWCPark2016}, i.e.\footnote{To make full use of the limited fronthaul capacity, the eRRH cooperation cluster should be optimized according to the fronthaul capacity and the cache status of each RRH. As this is not the focus of this work, we simply select the $N_{\mathrm{F}}$ eRRHs with the best channel qualities for the cooperation cluster.},
\begin{equation}\label{Cachenable53}
d_{f_{k},l,i}=
\begin{cases}
1,~~\text{if}~i\in\left\{i: \left|\left\{i': \left\|\mathbf{H}_{k,i'}\right\|_{\mathrm{F}}> \left\|\mathbf{H}_{k,i}\right\|_{\mathrm{F}}\right\}\right|< N_{\mathrm{F}}\right\} \text{and}~c_{f_{k},l,i}=0,\\
0 ,~~\text{otherwise} .
\end{cases}
\end{equation}

Let $R_{f,l}\leq\overline{S}$ be the data delivery rate of subfile $\left(f,l\right)$, so that $nR_{f,l}\leq n\overline{S}$ bits are transferred to the eRRH in the considered transmission interval~\cite{TWCPark2016}. The remaining $n\overline{S}-nR_{f,l}$ bits can be transferred in the following transmission interval. Thus, the fronthaul capacity constraint for eRRH $i$ is given by
\begin{equation}\label{Cachenable06}
\sum\limits_{f\in\mathcal{F}_{\mathrm{req}}}\sum\limits_{l\in\mathcal{L}}d_{f,l,i}R_{f,l}\leqslant C_{i}, \forall i\in\mathcal{K}_{\mathrm{R}}.
\end{equation}
Each eRRH $i$ performs precoding to generate the transmit signal $\mathbf{x}_{i}$ for subfile $\left(f,l\right)$, which has been cached or transferred via the fronthaul links to the eRRH. Then, signal $\mathbf{x}_{i}$ transmitted by eRRH $i$ can be expressed as:
\begin{equation}\label{Cachenable07}
\mathbf{x}_{i}=\mathbf{F}_{\mathrm{RF},i}\sum\limits_{f\in\mathcal{F}_{\mathrm{req}}}\sum\limits_{l\in\mathcal{L}}\left(1-\overline{c}_{f,l,i}\overline{d}_{f,l,i}\right)\mathbf{G}_{f,l,i}\mathbf{s}_{f,l},
\end{equation}
where $\mathbf{G}_{f,l,i}\in\mathbb{C}^{N_{\mathrm{tRF},i}\times d_{f,l}}$ is the digital precoding matrix at eRRH $i$ for baseband signal $\mathbf{s}_{f, l}\in\mathbb{C}^{d_{f,l}\times 1}$ encoding subfile $\left(f, l\right)$. $d_{f,l}$ denotes the number of spatial data streams used for subfile $\left(f,l\right)$. In this paper, we assume that $\sum\limits_{l\in\mathcal{L}}d_{f_{k},l}\leqslant \min\left(N_{\mathrm{uRF},k},N_{\mathrm{tRF}}\right)$, $\forall k\in\mathcal{K}_{\mathrm{U}}$, where $N_{\mathrm{tRF}}=\sum\limits_{i\in\mathcal{K}_{\mathrm{R}}}N_{\mathrm{tRF},i}$. $\mathbf{F}_{\mathrm{RF},i}\in\mathbb{C}^{N_{\mathrm{t},i}\times N_{\mathrm{tRF},i}}$ is the analog precoding matrix at eRRH $i$. Since $\mathbf{F}_{\mathrm{RF},i}$ is implemented via an analog phase shifter network, its elements are constrained to be constant-modulus. In what follows, we assume that each entry of $\mathbf{F}_{\mathrm{RF},i}$ has unit norm, i.e., $\left|\left[\mathbf{F}_{\mathrm{RF},i}\right]_{m_{i},n_{i}}\right|^{2}=1$, $m_{i}\in\mathcal{N}_{\mathrm{R},i}=\left\{1,\cdots,N_{\mathrm{t},i}\right\}$, and $n_{i}\in\mathcal{N}_{\mathrm{C},i}=\left\{1,\cdots,N_{\mathrm{tRF},i}\right\}$.

For convenience of presentation, the digital precoding matrix of all eRRHs for subfile $\left(f,l\right)$ and the channel matrix from all eRRHs to user $k$ are denoted by $\overline{\mathbf{U}}_{f,l}\triangleq\left[\overline{\mathbf{U}}_{f,l,1}^{H},\cdots,
\overline{\mathbf{U}}_{f,l,K_{\mathrm{R}}}^{H}\right]^{H}\in\mathbb{C}^{N_{\mathrm{tRF}}\times d_{f,l}}$ and $\mathbf{H}_{k}\triangleq\left[\mathbf{H}_{k,1},\cdots,\mathbf{H}_{k,K_{\mathrm{R}}}\right]\in\mathbb{C}^{N_{\mathrm{r},k}\times N_{\mathrm{tt}}}$, respectively, where $N_{\mathrm{tt}}=\sum\limits_{i\in\mathcal{K}_{\mathrm{R}}}N_{\mathrm{t},i}$ and $\overline{\mathbf{U}}_{f,l,i}=\left(1-\overline{c}_{f,l,i}\overline{d}_{f,l,i}\right)\mathbf{G}_{f,l,i}$. Then, the received signal $\mathbf{y}_{k}$ at user $k$ requesting file $f_{k}$ is
\begin{equation}\label{Cachenable08}
\mathbf{y}_{k}=\sum\limits_{l\in \mathcal{L}}\mathbf{H}_{k}\mathbf{F}_{\mathrm{RF}}\overline{\mathbf{U}}_{f_{k},l}\mathbf{s}_{f_{k},l}+
\sum\limits_{f\in\mathcal{F}_{\mathrm{req}}\setminus\left\{f_{k}\right\}}\sum\limits_{l\in \mathcal{L}}\mathbf{H}_{k}\mathbf{F}_{\mathrm{RF}}\overline{\mathbf{U}}_{f,l}\mathbf{s}_{f,l}+\mathbf{n}_{k},
\end{equation}
where $\mathbf{F}_{\mathrm{RF}}$ is the analog precoding super-matrix defined as:
\begin{equation}\label{Cachenable09}
\begin{split}
\mathbf{F}_{\mathrm{RF}}=&\left[\left(\mathbf{F}_{\mathrm{RF},1}\mathbf{P}_{1}^{H}\right)^{H},
\cdots,\left(\mathbf{F}_{\mathrm{RF},K_{\mathrm{R}}}\mathbf{P}_{K_{\mathrm{R}}}^{H}\right)^{H}\right]^{H}\\
=&\mathrm{diag}\left(\mathbf{F}_{\mathrm{RF},1},\cdots,\mathbf{F}_{\mathrm{RF},K_{\mathrm{R}}}\right),
\end{split}
\end{equation}
where permutation matrix $\mathbf{P}_{i}$ is defined as
$$\mathbf{P}_{i}=\left[\mathbf{0}_{N_{\mathrm{tRF},i}\times N_{\mathbf{P}_{i}}^{1}},\mathbf{I}_{N_{\mathrm{tRF},i}\times N_{\mathrm{tRF},i}},\mathbf{0}_{N_{\mathrm{tRF},i}\times N_{\mathbf{P}_{i}}^{2}}\right]^{T}$$
with $N_{\mathbf{P}_{i}}^{1}=\sum\limits_{j=1}^{i-1}N_{\mathrm{tRF},j}$ and $N_{\mathbf{P}_{i}}^{2}=N_{\mathrm{tRF}}-\sum\limits_{j=1}^{i}N_{\mathrm{tRF},j}$. Thus, the achieve data rate for subfile $\left(f_{k},l\right)$ at user $k$, in units of nats/s/Hz, is computed as
\begin{equation}\label{Cachenable10}
q_{k,l}\left(\mathbf{F}_{\mathrm{RF}},\mathcal{G}\right)\triangleq
\log\det\left(\mathbf{I}_{N_{\mathrm{r},k}\times N_{\mathrm{r},k}}+\overline{\mathbf{H}}_{k,f_{k},l}\overline{\mathbf{H}}_{k,f_{k},l}^{H}\mathbf{\Pi}_{k,l}^{-1}\right),
\end{equation}
where $\mathcal{G}\triangleq\left\{\mathbf{G}_{f,l,i}\right\}_{f\in \mathcal{F}_{\mathrm{req}},l\in \mathcal{L},i\in\mathcal{K}_{\mathrm{R}}}$, $\overline{\mathbf{H}}_{k,f,l}=\mathbf{H}_{k}\mathbf{F}_{\mathrm{RF}}\overline{\mathbf{U}}_{f,l}=\sum\limits_{i=1}^{K_{\mathrm{R}}}
\mathbf{H}_{k,i}\mathbf{F}_{\mathrm{RF},i}\overline{\mathbf{U}}_{f,l,i}$, and the interference-plus-noise covariance matrix $\mathbf{\Pi}_{k,l}$ is given by:
\begin{equation}\label{Cachenable11}
\mathbf{\Pi}_{k,l}=\sum\limits_{m\in \mathcal{L}\setminus\left\{l\right\}}\overline{\mathbf{H}}_{k,f_{k},m}\overline{\mathbf{H}}_{k,f_{k},m}^{H}
+\sum\limits_{f\in\mathcal{F}_{\mathrm{req}}\setminus\left\{f_{k}\right\}}\sum\limits_{\tau\in \mathcal{L}}\overline{\mathbf{H}}_{k,f,\tau}\overline{\mathbf{H}}_{k,f,\tau}^{H}
+\sigma_{k}^{2}\mathbf{I}_{N_{\mathrm{r},k}\times N_{\mathrm{r},k}}.
\end{equation}
In the following subsections, we investigate the design of the digital precoder $\mathcal{G}$ and the analog precoding super-matrix $\mathbf{F}_{\mathrm{RF}}$, where we take into account the fronthaul capacity constraint, the eRRH transmit power constraint, and the constant-modulus constraint on each element of the analog precoder.

\subsection*{A. Problem Formulation}
In this paper, our objective is to maximize the minimum user rate $R_{\mathrm{min}}$ defined as $R_{\mathrm{min}}\triangleq\min\limits_{f\in\mathcal{F}_{\mathrm{req}}} R_{f}$ subject to constraints on the fronthaul capacity, the file size, the eRRH transmit power, and the constant-modulus of each entry of the analog precoder, where $R_{f}=\sum\limits_{l\in\mathcal{L}}R_{f,l}$ represents the achievable delivery rate for file $f$. By maximizing the minimum user rate $R_{\mathrm{min}}$, the number of transmission intervals that are needed to deliver all files $\mathcal{F}_{\mathrm{req}}$ to the requesting users is minimized and fairness between the users is ensured~\cite{NetAh2014}. For a given cache status $c_{f,l,i}$ and fronthaul information transfer status $d_{f,l,i}$, $f\in\mathcal{F}$, $l\in\mathcal{L}$, $i\in\mathcal{K}_{\mathrm{R}}$, the design of the hybrid precoder for CeMm-RANs with HFIT is formulated as
\begin{subequations}\label{Cachenable12}
\begin{align}
&\max_{\mathbf{F}_{\mathrm{RF}},\mathcal{G}, \mathcal{R}}R_{\mathrm{min}},\label{Cachenable12a}\\
\mathrm{s.t.}~&\log\left(2\right)R_{f_{k},l}\leq q_{k,l}\left(\mathbf{F}_{\mathrm{RF}},\mathcal{G}\right), \forall k\in\mathcal{K}_{\mathrm{U}}, l\in\mathcal{L},\label{Cachenable12b}\\
&R_{\mathrm{min}}\leq\sum\limits_{l\in\mathcal{L}}R_{f,l}, \forall f\in\mathcal{F}_{\mathrm{req}},\label{Cachenable12c}\\
&R_{f,l}\leq \overline{S}, \forall f\in\mathcal{F}_{\mathrm{req}}, l\in\mathcal{L},\label{Cachenable12d}\\
&\sum\limits_{f\in\mathcal{F_{\mathrm{req}}}}\sum\limits_{l\in\mathcal{L}}d_{f,l,i}R_{f,l}\leqslant C_{i}, \forall i\in\mathcal{K}_{\mathrm{R}},\label{Cachenable12e}\\
&p_{i}\left(\mathbf{F}_{\mathrm{RF},i},\mathcal{G}\right)\leq P_{i}, \forall i\in\mathcal{K}_{\mathrm{R}},\label{Cachenable12f}\\
&\mathbf{F}_{\mathrm{RF},i}\in\mathcal{F}_{\mathrm{RF},i}, \forall i\in\mathcal{K}_{\mathrm{R}},\label{Cachenable12g}
\end{align}
\end{subequations}
where $\mathcal{F}_{\mathrm{RF},i}$ and $P_{i}$ denote the set of feasible analog precoders and the maximum transmit power for eRRH $i$, respectively. Furthermore, we define $\mathcal{R}=\left\{R_{\mathrm{min}}\right\}\cup\left\{R_{f,l}\right\}_{f\in\mathcal{F}_{\mathrm{req}}, l\in\mathcal{L}}$, and $p_{i}\left(\mathbf{F}_{\mathrm{RF},i},\mathcal{G}\right)$ is given by
\begin{equation}\label{Cachenable13}
p_{i}\left(\mathbf{F}_{\mathrm{RF},i},\mathcal{G}\right)\triangleq\sum\limits_{f\in\mathcal{F}_{\mathrm{req}}}
\sum\limits_{l\in\mathcal{L}}\mathrm{tr}\left(\mathbf{F}_{\mathrm{RF},i}\mathbf{P}_{i}^{H}\overline{\mathbf{U}}_{f,l}\overline{\mathbf{U}}_{f,l}^{H}
\mathbf{P}_{i}\mathbf{F}_{\mathrm{RF},i}^{H}\right)
=\sum\limits_{f\in\mathcal{F}_{\mathrm{req}}}
\sum\limits_{l\in\mathcal{L}}\mathrm{tr}\left(\mathbf{F}_{\mathrm{RF},i}\overline{\mathbf{U}}_{f,l,i}\overline{\mathbf{U}}_{f,l,i}^{H}
\mathbf{F}_{\mathrm{RF},i}^{H}\right).
\end{equation}
In problem~\eqref{Cachenable12}, constraint~\eqref{Cachenable12d} ensures that the data delivery rate of subfile $\left(f,l\right)$ does not exceed the normalized subfile size $\overline{S}$. The limited fronthaul capacity $C_{i}$, $i\in\mathcal{K}_{\mathrm{R}}$, constrains the rate on each fronthaul link in~\eqref{Cachenable12e}. Constraint~\eqref{Cachenable12f} limits the maximum allowable transmit power of eRRH $i$. The optimal value of~\eqref{Cachenable12} is mainly limited by constraints~\eqref{Cachenable12b},~\eqref{Cachenable12d},~\eqref{Cachenable12e}, and~\eqref{Cachenable12f}. Problem~\eqref{Cachenable12} includes the non-convex rate constraint in~\eqref{Cachenable12b}, the constant-modulus constraint on the entries of the analog precoder in~\eqref{Cachenable12g}, and the strong coupling between the analog precoders and the digital precoders. Thus, problem~\eqref{Cachenable12} is in general difficult to solve globally. Therefore, first, we resort to a convex approximation approach to transform problem~\eqref{Cachenable12} into a tractable form, and then we develop an effective approach to obtain a solution of problem~\eqref{Cachenable12}.
\subsection*{B. Optimization of Digital Precoder}
To avoid the coupling between the analog precoding matrices and the digital precoding matrices, we first optimize the digital precoding matrices for given analog precoding matrices. For fixed analog precoders, problem~\eqref{Cachenable12} can be reformulated as:
\begin{equation}\label{Cachenable54}
\max_{\mathcal{G},\mathcal{R}}R_{\mathrm{min}}~	\mathrm{s.t.}~\eqref{Cachenable12b},\eqref{Cachenable12c},\eqref{Cachenable12d},\eqref{Cachenable12e},\eqref{Cachenable12f}.
\end{equation}
Note that the difficulty in solving problem~\eqref{Cachenable54} lies in constraint~\eqref{Cachenable12b}, as the achievable data rate $q_{k,l}\left(\mathbf{F}_{\mathrm{RF}},\mathcal{G}\right)$ is non-convex. To overcome this difficulty, in the sequel, we resort to approximating the achievable data rate $q_{k,l}\left(\mathbf{F}_{\mathrm{RF}},\mathcal{G}\right)$ by a convex lower bound.

We first note that the achievable data rate $q_{k,l}\left(\mathbf{F}_{\mathrm{RF}},\mathcal{G}\right)$ of subfile $\left(f_{k},l\right)$ can be expressed as a function of the error covariance matrix after minimum-mean-square-error (MMSE) receive filtering~\cite{TWCChris2008}. Let $\mathbf{\Xi}_{k,l}\in\mathbb{C}^{N_{\mathrm{r},k}\times d_{f_{k},l}}$ be a linear receiver applied at user $k$ for recovering subfile $\left(f_{k},l\right)$. Thus, the mean-square-error (MSE) matrix for recovering subfile $\left(f_{k},l\right)$ is calculated as
\begin{equation}\label{Cachenable14}
\begin{split}
\mathbf{E}_{k,l}
&=\mathbb{E}\left[\left(\mathbf{\Xi}_{k,l}^{H}\mathbf{y}_{k}-\mathbf{s}_{f_{k},l}\right)
\left(\mathbf{\Xi}_{k,l}^{H}\mathbf{y}_{k}-\mathbf{s}_{f_{k},l}\right)^{H}\right]\\
&=\mathbf{I}_{d_{f_{k},l}\times d_{f_{k},l}}-\overline{\mathbf{H}}_{k,f_{k},l}^{H}\mathbf{\Xi}_{k,l}-\mathbf{\Xi}_{k,l}^{H}\overline{\mathbf{H}}_{k,f_{k},l}+
\mathbf{\Xi}_{k,l}^{H}\mathbf{\Lambda}_{k,l}\mathbf{\Xi}_{k,l},
\end{split}
\end{equation}
where $\mathbf{\Lambda}_{k,l}$ is given by
\begin{equation}\label{Cachenable15}
\mathbf{\Lambda}_{k,l}=\sum\limits_{f\in\mathcal{F}_{\mathrm{req}}}\sum\limits_{\tau\in \mathcal{L}}\overline{\mathbf{H}}_{k,f,\tau}\overline{\mathbf{H}}_{k,f,\tau}^{H}+\sigma_{k}^{2}\mathbf{I}_{N_{\mathrm{r},k}\times N_{\mathrm{r},k}}.
\end{equation}
Note that for fixed analog precoder and receive filter,~\eqref{Cachenable14} is a convex function with respect to (w.r.t.) the digital precoder $\mathcal{G}$. Thus, we can exploit this feature to convexify the achievable data rate $q_{k,l}\left(\mathbf{F}_{\mathrm{RF}},\mathcal{G}\right)$. According to~\eqref{Cachenable14}, the MMSE-receive filter at user $k$ can be obtained as
\begin{equation}\label{Cachenable16}
\mathbf{\Xi}_{k,l}^{\mathrm{MMSE}}=\mathbf{\Lambda}_{k,l}^{-1}\overline{\mathbf{H}}_{k,f_{k},l}.
\end{equation}
The MSE-matrix at user $k$ for the MMSE-receive filter can be written as:
\begin{equation}\label{Cachenable17}
\begin{split}
\mathbf{E}_{k,l}^{\mathrm{MMSE}}
&=\mathbf{I}_{d_{f_{k},l}\times d_{f_{k},l}}-\overline{\mathbf{H}}_{k,f_{k},l}^{H}\mathbf{\Lambda}_{k,l}^{-1}\overline{\mathbf{H}}_{k,f_{k},l}
=\left(\mathbf{I}_{d_{f_{k},l}\times d_{f_{k},l}}+\overline{\mathbf{H}}_{k,f_{k},l}^{H}\mathbf{\Pi}_{k,l}^{-1}\overline{\mathbf{H}}_{k,f_{k},l}\right)^{-1}.
\end{split}
\end{equation}
Combining~\eqref{Cachenable10} and~\eqref{Cachenable17}, we have $q_{k,l}\left(\mathbf{F}_{\mathrm{RF}},\mathcal{G}\right)=\log\det\left(\left(\mathbf{E}_{k,l}^{\mathrm{MMSE}}\right)^{-1}\right)$.
The authors of~\cite{TWCChris2008} have shown that the achievable data rate $q_{k,l}\left(\mathbf{F}_{\mathrm{RF}},\mathcal{G}\right)$ for subfile $\left(f_{k},l\right)$ can be expressed as
\begin{equation}\label{Cachenable18}
q_{k,l}\left(\mathbf{F}_{\mathrm{RF}},\mathcal{G}\right)\triangleq\max\limits_{\mathbf{\Upsilon}_{k,l},\mathbf{\Xi}_{k,l}}
\left(\log\det\left(\mathbf{\Upsilon}_{k,l}\right)-\mathrm{tr}\left(\mathbf{\Upsilon}_{k,l}\mathbf{E}_{k,l}\right)
+d_{f_{k},l}\right),
\end{equation}
where $\mathbf{\Upsilon}_{k,l}\succeq \bm{0}$ is a weight matrix. For simplicity, define function $\widetilde{q}_{k,l}\left(\mathbf{F}_{\mathrm{RF}},\mathcal{G}, \mathcal{V}, \mathcal{X}\right)$ as
\begin{equation}\label{Cachenable21}
\widetilde{q}_{k,l}\left(\mathbf{F}_{\mathrm{RF}},\mathcal{G}, \mathcal{V}, \mathcal{X}\right)\triangleq
\log\det\left(\mathbf{\Upsilon}_{k,l}\right)-\mathrm{tr}\left(\mathbf{\Upsilon}_{k,l}\mathbf{E}_{k,l}\right)
+d_{f_{k},l}
\end{equation}
with $\mathcal{V}=\left\{\mathbf{\Upsilon}_{k,l}\right\}_{k\in\mathcal{K}_{\mathrm{U}}, l\in\mathcal{L}}$ and $\mathcal{X}=\left\{\mathbf{\Xi}_{k,l}\right\}_{k\in\mathcal{K}_{\mathrm{U}}, l\in\mathcal{L}}$. According to~\eqref{Cachenable18}, we have $\widetilde{q}_{k,l}\left(\mathbf{F}_{\mathrm{RF}},\mathcal{G}, \mathcal{V}, \mathcal{X}\right)$ $\leq q_{k,l}\left(\mathbf{F}_{\mathrm{RF}},\mathcal{G}\right)$. Note that function $\widetilde{q}_{k,l}\left(\mathbf{F}_{\mathrm{RF}},\mathcal{G}, \mathcal{V}, \mathcal{X}\right)$ is convex w.r.t. each individual optimization variable $\mathcal{G}$, $\mathcal{V}$, and $\mathcal{X}$, but is not jointly convex in these variables. Based on this observation, we replace the achievable data rate $q_{k,l}\left(\mathbf{F}_{\mathrm{RF}},\mathcal{G}\right)$ in constraint~\eqref{Cachenable12b} by $\widetilde{q}_{k,l}\left(\mathbf{F}_{\mathrm{RF}},\mathcal{G}, \mathcal{V}, \mathcal{X}\right)$ and solve the following problem instead of problem~\eqref{Cachenable54}:
\begin{subequations}\label{Cachenable20}
\begin{align}
&\max_{\mathcal{G},\mathcal{R}, \mathcal{V},\mathcal{X}}R_{\mathrm{min}},\label{Cachenable20a}\\
\mathrm{s.t.}~&\log\left(2\right)R_{f_{k},l}\leq \widetilde{q}_{k,l}\left(\mathbf{F}_{\mathrm{RF}},\mathcal{G}, \mathcal{V}, \mathcal{X}\right), \forall k\in\mathcal{K}_{\mathrm{U}}, l\in\mathcal{L},\label{Cachenable20b}\\
&\eqref{Cachenable12c},\eqref{Cachenable12d},\eqref{Cachenable12e},\eqref{Cachenable12f}.\label{Cachenable20c}
\end{align}
\end{subequations}
Problem~\eqref{Cachenable20} is still a non-convex problem due to the coupling between the optimization variables in~\eqref{Cachenable20b}. However, problem~\eqref{Cachenable20} is a convex optimization problem w.r.t. each individual optimization variable $\mathcal{G}$, $\mathcal{R}$, $\mathcal{V}$, and $\mathcal{X}$. Therefore, the block coordinate ascent method~\cite{TWCChris2008} is adopted to solve problem~\eqref{Cachenable20}. In particular, we first solve problem~\eqref{Cachenable20} w.r.t. variables $\mathcal{V}$ and $\mathcal{X}$, respectively, for fixed $\mathcal{G}$ and $\mathcal{R}$. Then, we solve problem~\eqref{Cachenable20} w.r.t. variables $\mathcal{G}$ and $\mathcal{R}$ for fixed $\mathcal{V}$ and $\mathcal{X}$. For given $\mathcal{G}$ and $\mathcal{R}$, the optimal $\mathbf{\Xi}_{k,l}$ of problem~\eqref{Cachenable20} is the MMSE-receive filter in \eqref{Cachenable16} and the optimal $\mathbf{\Upsilon}_{k,l}$ of problem~\eqref{Cachenable20} is given by:
\begin{equation}\label{Cachenable19}
\mathbf{\Upsilon}_{k,l}^{\mathrm{opt}}=\left(\mathbf{E}_{k,l}^{\mathrm{MMSE}}\right)^{-1}.
\end{equation}
For given $\mathcal{X}$ and $\mathcal{V}$, the optimal $\mathcal{G}$ and $\mathcal{R}$ in~\eqref{Cachenable20} can be found by solving the following problem
\begin{equation}\label{Cachenable22}
\max_{\mathcal{G}, \mathcal{R}}R_{\mathrm{min}}~
\mathrm{s.t.}~\eqref{Cachenable12c},\eqref{Cachenable12d},\eqref{Cachenable12e},\eqref{Cachenable12f},\eqref{Cachenable20b}.
\end{equation}
Problem~\eqref{Cachenable22} is a convex optimization problem that can be solved by convex program solvers such as CVX~\cite{{AvialGrant2015}}. The procedure used for solving problem~\eqref{Cachenable20} is summarized in Algorithm~\ref{CachenableAlg01} where $t$ denotes the number of iterations and $\epsilon$ denotes a predefined stopping criterion.
\begin{algorithm}[t]
\caption{Solution of problem~\eqref{Cachenable20}}\label{CachenableAlg01}
\begin{algorithmic}[1]
\STATE Analog  precoding super-matrix $\mathbf{F}_{\mathrm{RF}}$ is given.\label{CachenableAlg0101}
\STATE Set $t=0$ and initialize $\mathcal{G}^{\left(t\right)}$ and $\mathcal{R}^{\left(t\right)}$ such that constraints \eqref{Cachenable12b} to \eqref{Cachenable12f} are satisfied.\label{CachenableAlg0102}
\STATE Compute the MMSE-receive filters $\mathcal{X}^{\left(t+1\right)}$ for the given $\mathcal{G}^{\left(t\right)}$ and $\mathbf{F}_{\mathrm{RF}}$ using~\eqref{Cachenable16}.\label{CachenableAlg0103}
\STATE Compute the weight matrices $\mathcal{V}^{\left(t+1\right)}$ for the given $\mathcal{G}^{\left(t\right)}$ and $\mathbf{F}_{\mathrm{RF}}$ using~\eqref{Cachenable19}.\label{CachenableAlg0104}
\STATE Solve problem~\eqref{Cachenable22} to obtain $\mathcal{G}^{\left(t+1\right)}$ and $\mathcal{R}^{\left(t+1\right)}$ for the given $\mathbf{F}_{\mathrm{RF}}$, $\mathcal{X}^{\left(t+1\right)}$, and  $\mathcal{V}^{\left(t+1\right)}$.\label{CachenableAlg0105}
\STATE If $\left| R_{\mathrm{min}}^{\left(t+1\right)}-R_{\mathrm{min}}^{\left(t\right)}\right|\leqslant\epsilon$, then stop the iteration and output $\mathcal{G}^{\left(t+1\right)}$ and $\mathcal{R}^{\left(t+1\right)}$. Otherwise, set $t\leftarrow t+1$ and go to Step~\ref{CachenableAlg0103}.\label{CachenableAlg0106}
\end{algorithmic}
\end{algorithm}

From Algorithm~\ref{CachenableAlg01}, we observe that each update of the MMSE-receive filters $\mathcal{X}$ and the weight matrices $\mathcal{V}$ in Steps~\ref{CachenableAlg0103} and \ref{CachenableAlg0104}, respectively, maximizes function $ \widetilde{q}_{k,l}\left(\mathbf{F}_{\mathrm{RF}},\mathcal{G}, \mathcal{V}, \mathcal{X}\right)$, i.e., the right hand side of constraint~\eqref{Cachenable20b}, without affecting the other constraints in problem~\eqref{Cachenable20}. Hence, the update of the MMSE-receive filters $\mathcal{X}$ and the weight matrices $\mathcal{V}$ does not change the objective value, but may increase the feasible set for improving the minimum user rate. On the other hand, the update of the optimization variables $\mathcal{G}$ and $\mathcal{R}$ in Step \ref{CachenableAlg0105} of Algorithm~\ref{CachenableAlg01} maximizes the objective value of problem~\eqref{Cachenable20} for the given $\mathcal{X}$ and $\mathcal{V}$. Therefore, we have the following sequence as the iterations $t$ proceed
$$R_{\mathrm{min}}^{\left(0\right)}\leq R_{\mathrm{min}}^{\left(1\right)}\leq\cdots\leq R_{\mathrm{min}}^{\left(t\right)}\leq R_{\mathrm{min}}^{\left(t+1\right)}\leq\cdots,$$
i.e., Algorithm~\ref{CachenableAlg01} generates a monotonically non-decreasing sequence of objective values. In addition, the objective  function of problem~\eqref{Cachenable20} is upper bounded due to the limited transmit power of each eRRH. Therefore, Algorithm~\ref{CachenableAlg01} is guaranteed to converge~\cite{Bibby1974}. Using the same arguments as in~\cite[Theorem 1]{OperalMarks1977} and~\cite[Theorem 3]{TSPShi2011}, we can prove that Algorithm~\ref{CachenableAlg01} converges to a stationary point of problem~\eqref{Cachenable20}.

Algorithm~\ref{CachenableAlg01} involves matrix multiplications, matrix inversions, and solving a convex problem.  The computational complexity of Steps~\ref{CachenableAlg0103} and~\ref{CachenableAlg0104} of Algorithm~\ref{CachenableAlg01} is $K_{\mathrm{U}}L\mathit{O}\left(N_{v,1}\right)$, where $\mathit{O}\left(\cdot\right)$ stands for the big-O notation, and $N_{v,1}=N_{\mathrm{rt}}N_{\mathrm{tt}}N_{\mathrm{RF}}L+K_{\mathrm{U}}N_{\mathrm{uRF}}^{2.736}$ with $N_{\mathrm{rt}}=\sum\limits_{k\in\mathcal{K}_{\mathrm{U}}}N_{r,k}$ and $N_{\mathrm{uRF}}=\max\limits_{k\in\mathcal{K}_{\mathrm{U}}}N_{\mathrm{uRF},k}$~\cite{MatrixHorn1986}. In Step~\ref{CachenableAlg0105} of Algorithm~\ref{CachenableAlg01}, a convex optimization problem is solved, which can be efficiently implemented by a primal-dual interior point method with an approximate complexity of $\mathit{O}\left(\phi_{1}\right)$, where $\phi_{1}=\left(LK_{\mathrm{U}}\left(N_{\mathrm{tRF}}d_{\mathrm{m}}K_{\mathrm{R}}+1\right)\right)^{3.5}$ and $d_{\mathrm{m}}=\max\limits_{f\in\mathcal{F}_{\mathrm{req}},l\in\mathcal{L}}d_{f,l}$~\cite{MathPotra2000}. Suppose that Algorithm~\ref{CachenableAlg01} needs $\kappa_{1}$ iterations to converge. Then, the overall computational complexity of Algorithm~\ref{CachenableAlg01} is $\mathit{O}\left(\kappa_{1}\left(K_{\mathrm{U}}LN_{v,1}+\phi_{1}\right)\right)$.

\subsection*{C. Optimization of Analog Precoder}
In this subsection, we focus on optimizing the analog precoders by solving problem~\eqref{Cachenable12} for fixed digital precoders. In particular,  we propose to iteratively solve the following problem for given digital precoding matrices $\mathcal{G}$:
\begin{equation}\label{Cachenable24}
\max_{\mathbf{F}_{\mathrm{RF}},\mathcal{R}, \mathcal{V},\mathcal{X}}R_{\mathrm{min}}~
\mathrm{s.t.}~\eqref{Cachenable12c},\eqref{Cachenable12d},\eqref{Cachenable12e},\eqref{Cachenable12f},\eqref{Cachenable12g},\eqref{Cachenable20b}.
\end{equation}
The optimal solutions of $\mathcal{V}$ and $\mathcal{X}$ are still given by~\eqref{Cachenable16} and~\eqref{Cachenable19}, respectively. The main difficulty in solving problem~\eqref{Cachenable24} is the constant-modulus constraint, which makes the optimization problem more challenging than that for the digital precoders. Therefore, in the following, we iteratively convexify the constant-modulus constraint such that a tractable form is obtained.

For ease of  presentation, we denote the analog precoding matrix in the $t$-th iteration by $\mathbf{F}_{\mathrm{RF},i}^{\left(t\right)}$ and further assume that the initial matix $\mathbf{F}_{\mathrm{RF},i}^{\left(0\right)}$ is given. Also, we denote the phase of the $\left(m_{i},n_{i}\right)$-th entry of $\mathbf{F}_{\mathrm{RF},i}^{\left(t\right)}$ as $\varphi_{m_{i},n_{i}}^{\left(t\right)}$. In the sequel, we update $\mathbf{F}_{\mathrm{RF},i}^{\left(t+1\right)}$ using a local  search in a small vicinity of $\mathbf{F}_{\mathrm{RF},i}^{\left(t\right)}$. The $\left(m_{i},n_{i}\right)$-th entry of matrix $\mathbf{F}_{\mathrm{RF},i}^{\left(t+1\right)}$ is updated as $e^{j\varphi_{m_{i},n_{i}}^{\left(t+1\right)}}=e^{j\left(\varphi_{m_{i},n_{i}}^{\left(t\right)}+\delta_{m_{i},n_{i}}^{\left(t\right)}\right)}$ with $\delta_{m_{i},n_{i}}^{\left(t\right)}$ being the phase increment of the $\left(m_{i},n_{i}\right)$-th entry of $\mathbf{F}_{\mathrm{RF},i}^{\left(t\right)}$, i.e.,
\begin{equation}\label{Cachenable25}
\left[\mathbf{F}_{\mathrm{RF},i}^{\left(t+1\right)}\right]_{m_{i},n_{i}}=e^{j\left(\varphi_{m_{i},n_{i}}^{\left(t\right)}+\delta_{m_{i},n_{i}}^{\left(t\right)}\right)}.
\end{equation}
To obtain a tractable form of~\eqref{Cachenable24} based on the expression for each updated entry of the analog precoder in~\eqref{Cachenable25}, exploiting a Taylor series expansion, $\mathbf{F}_{\mathrm{RF},i}^{\left(t+1\right)}$ is approximated as
\begin{equation}\label{Cachenable26}
\begin{split}
\mathbf{F}_{\mathrm{RF},i}^{\left(t+1\right)}\approx\mathbf{F}_{\mathrm{RF},i}^{\left(t\right)}+\widehat{\mathbf{F}}_{\mathrm{RF},i}^{\left(t\right)}\circ j\mathbf{F}_{\mathrm{RF},i}^{\left(t\right)},
\end{split}
\end{equation}
where the $\left(m_{i},n_{i}\right)$-th entry of matrix $\widehat{\mathbf{F}}_{\mathrm{RF},i}^{\left(t\right)}$ is $\delta_{m_{i},n_{i}}^{\left(t\right)}$.  Note that the approximation in~\eqref{Cachenable26} is inferred from $e^{j\delta_{m_{i},n_{i}}^{\left(t\right)}}\approx1+j\delta_{m_{i},n_{i}}^{\left(t\right)}$ which holds as long as $\delta_{m_{i},n_{i}}^{\left(t\right)}$ is sufficiently small, e.g., $\left|\delta_{m_{i},n_{i}}^{\left(t\right)}\right|\leqslant 0.1$.  Unfortunately, the formulation of~\eqref{Cachenable26} in terms of a Hadamard product is not compatible with many convex problem solvers. Therefore, we need to rewrite~\eqref{Cachenable26} in a more appropriate form such that convex programm solvers can be used to optimize the analog precoders. Note that $\mathbf{\Upsilon}_{k,l}\succeq \bm{0}$ and let $\mathbf{\Upsilon}_{k,l}=\overline{\mathbf{\Upsilon}}_{k,l}\overline{\mathbf{\Upsilon}}_{k,l}^{H}$, then the term $\mathrm{tr}\left(\mathbf{\Upsilon}_{k,l}\mathbf{E}_{k,l}\right)$ in~\eqref{Cachenable21} in the $\left(t+1\right)$-th iteration can be rewritten as:
\begin{equation}\label{Cachenable27}
\begin{split}
&\mathrm{tr}\left(\mathbf{\Upsilon}_{k,l}^{\left(t\right)}\mathbf{E}_{k,l}^{\left(t+1\right)}\right)
=\left\|\overline{\mathbf{\Upsilon}}_{k,l}^{\left(t\right)}-\overline{\mathbf{U}}_{f_{k},l}^{H}\left(\mathbf{F}_{\mathrm{RF}}^{\left(t\right)}\right)^{H}
\mathbf{H}_{k}^{H}\mathbf{\Xi}_{k,l}^{\left(t\right)}\overline{\mathbf{\Upsilon}}_{k,l}^{\left(t\right)}
-\overline{\mathbf{U}}_{f_{k},l}^{H}\left(\widetilde{\mathbf{F}}_{\mathrm{RF}}^{\left(t\right)}\right)^{H}
\mathbf{H}_{k}^{H}\mathbf{\Xi}_{k,l}^{\left(t\right)}\overline{\mathbf{\Upsilon}}_{k,l}^{\left(t\right)}\right\|_{\mathrm{F}}^{2}\\
&+\sum\limits_{m\in\mathcal{L}\setminus\left\{l\right\}}\left\|\overline{\mathbf{U}}_{f_{k},m}^{H}\left(\mathbf{F}_{\mathrm{RF}}^{\left(t\right)}\right)^{H}
\mathbf{H}_{k}^{H}\mathbf{\Xi}_{k,l}^{\left(t\right)}\overline{\mathbf{\Upsilon}}_{k,l}^{\left(t\right)}
+\overline{\mathbf{U}}_{f_{k},m}^{H}\left(\widetilde{\mathbf{F}}_{\mathrm{RF}}^{\left(t\right)}\right)^{H}
\mathbf{H}_{k}^{H}\mathbf{\Xi}_{k,l}^{\left(t\right)}\overline{\mathbf{\Upsilon}}_{k,l}^{\left(t\right)}\right\|_{\mathrm{F}}^{2}\\
&+\sum\limits_{f\in\mathcal{F}_{\mathrm{req}}\setminus\left\{f_{k}\right\}}\sum\limits_{\tau\in \mathcal{L}}\left\|\overline{\mathbf{U}}_{f,\tau}^{H}\left(\mathbf{F}_{\mathrm{RF}}^{\left(t\right)}\right)^{H}\mathbf{H}_{k}^{H}\mathbf{\Xi}_{k,l}^{\left(t\right)}\overline{\mathbf{\Upsilon}}_{k,l}^{\left(t\right)}+\overline{\mathbf{U}}_{f,\tau}^{H}\left(\widetilde{\mathbf{F}}_{\mathrm{RF}}^{\left(t\right)}\right)^{H}\mathbf{H}_{k}^{H}\mathbf{\Xi}_{k,l}^{\left(t\right)}\overline{\mathbf{\Upsilon}}_{k,l}^{\left(t\right)}\right\|_{\mathrm{F}}^{2}\\
&+\mathrm{tr}\left(\sigma_{k}^{2}\left(\mathbf{\Xi}_{k,l}^{\left(t\right)}\right)^{H}\mathbf{\Xi}_{k,l}^{\left(t\right)}\mathbf{\Upsilon}_{k,l}^{\left(t\right)}\right).
\end{split}
\end{equation}
where $\widetilde{\mathbf{F}}_{\mathrm{RF}}^{\left(t\right)}=\widehat{\mathbf{F}}_{\mathrm{RF}}^{\left(t\right)}\circ j\mathbf{F}_{\mathrm{RF}}^{\left(t\right)}$ and $\widehat{\mathbf{F}}_{\mathrm{RF}}=\mathrm{diag}\left(\widehat{\mathbf{F}}_{\mathrm{RF},1},\cdots,\widehat{\mathbf{F}}_{\mathrm{RF},K_{\mathrm{R}}}\right)$.
Exploiting $\mathrm{vec}\left(\mathbf{A}\circ\mathbf{B}\right)=\mathrm{vec}\left(\mathbf{A}\right)\circ\mathrm{vec}\left(\mathbf{B}\right)=
\mathrm{diag}\left(\mathrm{vec}\left(\mathbf{A}\right)\right)\mathrm{vec}\left(\mathbf{B}\right)$ and $\mathrm{vec}\left(\mathbf{A}\mathbf{B}\mathbf{C}\right)=\left(\mathbf{C}^{T}\circledast\mathbf{A}\right)\mathrm{vec}\left(\mathbf{B}\right)$ ~\cite{MatrixHorn1986}, we can rewrite~\eqref{Cachenable27} as follows:
\begin{equation}\label{Cachenable28}
\begin{split}
\mathrm{tr}\left(\mathbf{\Upsilon}_{k,l}^{\left(t\right)}\mathbf{E}_{k,l}^{\left(t+1\right)}\right)
&=\left\|\mathbf{b}_{k,l,f_{k},l}^{\left(t\right)}-\widetilde{\mathbf{c}}_{k,l,f_{k},l}^{\left(t\right)}\right\|_{\mathrm{F}}^{2}
+\sum\limits_{m\in\mathcal{L}\setminus\left\{l\right\}}\left\|\mathbf{c}_{k,l,f_{k},m}^{\left(t\right)}+\widetilde{\mathbf{c}}_{k,l,f_{k},m}^{\left(t\right)}\right\|_{\mathrm{F}}^{2}\\
&+\sum\limits_{f\in\mathcal{F}_{\mathrm{req}}\setminus\left\{f_{k}\right\}}\sum\limits_{\tau\in \mathcal{L}}\left\|\mathbf{c}_{k,l,f,\tau}^{\left(t\right)}+\widetilde{\mathbf{c}}_{k,l,f,\tau}^{\left(t\right)}\right\|_{\mathrm{F}}^{2}+\mathrm{tr}\left(\sigma_{k}^{2}\left(\mathbf{\Xi}_{k,l}^{\left(t\right)}\right)^{H}\mathbf{\Xi}_{k,l}^{\left(t\right)}\mathbf{\Upsilon}_{k,l}^{\left(t\right)}\right),
\end{split}
\end{equation}
where $\mathbf{a}_{k,l}^{\left(t\right)}=\mathrm{vec}\left(\overline{\mathbf{\Upsilon}}_{k,l}^{\left(t\right)}\right)$, $\mathbf{b}_{k,l,f,\tau}^{\left(t\right)}=\mathbf{a}_{k,l}^{\left(t\right)}-\mathbf{c}_{k,l,f,\tau}^{\left(t\right)}$, $\mathbf{c}_{k,l,f,\tau}^{\left(t\right)}=\mathbf{A}_{k,l,f,\tau}^{\left(t\right)}\mathrm{vec}\left(\left(\mathbf{F}_{\mathrm{RF}}
^{\left(t\right)}\right)^{H}\right)$ with $\mathbf{A}_{k,l,f,\tau}^{\left(t\right)}=\left(\left(\overline{\mathbf{\Upsilon}}_{k,l}^{\left(t\right)}
\right)^{T}\left(\mathbf{\Xi}_{k,l}^{\left(t\right)}\right)^{T}\mathbf{H}_{k}^{*}\right)\circledast\overline{\mathbf{U}}_{f,\tau}^{H}$, and
\begin{equation}
\widetilde{\mathbf{c}}_{k,l,f,\tau}^{\left(t\right)}=\mathbf{A}_{k,l,f,\tau}^{\left(t\right)}\mathrm{vec}\left(\left(\widetilde{\mathbf{F}}_{\mathrm{RF}}^{\left(t\right)}\right)^{H}\right)=\mathbf{A}_{k,l,f,\tau}^{\left(t\right)}\mathrm{diag}\left(\mathrm{vec}\left(\left(\widehat{\mathbf{F}}_{\mathrm{RF}}^{\left(t\right)}\right)^{H}\right)\right)\mathrm{vec}\left(\left(j\mathbf{F}_{\mathrm{RF}}^{\left(t\right)}\right)^{H}\right).
\end{equation}
Thus, analog precoding matrix $\mathbf{F}_{\mathrm{RF},i}^{\left(t+1\right)}$, $i\in\mathcal{K}_{\mathrm{R}}$, can be  updated by solving the following problem:
\begin{subequations}\label{Cachenable29}
\begin{align}
&\max_{\mathcal{\delta}^{\left(t\right)},\mathcal{R}, \mathcal{V},\mathcal{X}}R_{\mathrm{min}},\label{Cachenable29a}\\
\mathrm{s.t.}~&\log\left(2\right)R_{f_{k},l}\leq \widehat{q}_{k,l}^{\left(t\right)}\left(\mathcal{\delta}^{\left(t\right)}, \mathcal{V}, \mathcal{X}\right), \forall k\in\mathcal{K}_{\mathrm{U}}, l\in\mathcal{L},\label{Cachenable29b}\\
&\eqref{Cachenable12c},\eqref{Cachenable12d}, \eqref{Cachenable12e},\label{Cachenable29c}\\
&\widehat{p}_{i}^{\left(t\right)}\left(\mathcal{\delta}^{\left(t\right)}\right)\leq P_{i}, \forall i\in\mathcal{K}_{\mathrm{R}},\label{Cachenable29d}\\
&\left|\delta_{m_{i},n_{i}}^{\left(t\right)}\right|\leqslant\varepsilon^{\left(t\right)},\forall i\in\mathcal{K}_{\mathrm{R}}, m_{i}\in\mathcal{N}_{\mathrm{R}, i}, n_{i}\in\mathcal{N}_{\mathrm{C}, i},\label{Cachenable29e}
\end{align}
\end{subequations}
where $\mathcal{\delta}^{\left(t\right)}=\left\{\delta_{m_{i},n_{i}}^{\left(t\right)}\right\}_{\forall i\in\mathcal{K}_{\mathrm{R}}, m_{i}\in\mathcal{N}_{\mathrm{R}, i}, n_{i}\in\mathcal{N}_{\mathrm{C}, i}}$ and $\widehat{q}_{k,l}^{\left(t\right)}\left(\mathcal{\delta}, \mathcal{V}, \mathcal{X}\right)$ is defined as
\begin{equation}\label{Cachenable30}
\widehat{q}_{k,l}^{\left(t\right)}\left(\mathcal{\delta}^{\left(t\right)}, \mathcal{V}, \mathcal{X}\right)\triangleq
\log\det\left(\mathbf{\Upsilon}_{k,l}^{\left(t\right)}\right)-\mathrm{tr}\left(\mathbf{\Upsilon}_{k,l}^{\left(t\right)}\mathbf{E}_{k,l}^{\left(t+1\right)}\right)
+d_{f_{k},l},
\end{equation}
In~\eqref{Cachenable29}, $\varepsilon^{\left(t\right)}>0$ is sufficiently small such that $e^{j\delta_{m_{i},n_{i}}^{\left(t\right)}}\approx1+j\delta_{m_{i},n_{i}}^{\left(t\right)}$ holds\footnote{A discussion on suitable choices of $\varepsilon^{\left(t\right)}$ can be found in~\cite{TSPNi2017}.} and $\widehat{p}_{i}^{\left(t\right)}\left(\mathcal{\delta}^{\left(t\right)}\right)$ is defined as
\begin{equation}\label{Cachenable31}
\begin{split}
&\widehat{p}_{i}^{\left(t\right)}\left(\mathcal{\delta}^{\left(t\right)}\right)\triangleq\sum\limits_{f\in\mathcal{F}_{\mathrm{req}}}
\sum\limits_{l\in\mathcal{L}}\left\|\left(\mathbf{F}_{\mathrm{RF},i}^{\left(t\right)}+\widehat{\mathbf{F}}_{\mathrm{RF},i}^{\left(t\right)}\circ j\mathbf{F}_{\mathrm{RF},i}^{\left(t\right)}\right)\overline{\mathbf{U}}_{f,l,i}\right\|_{\mathrm{F}}^{2}\\
&=\sum\limits_{f\in\mathcal{F}_{\mathrm{req}}}
\sum\limits_{l\in\mathcal{L}}\left\|\mathrm{vec}\left(\mathbf{F}_{\mathrm{RF},i}^{\left(t\right)}\overline{\mathbf{U}}_{f,l,i}\right)+
\left(\overline{\mathbf{U}}_{f,l,i}^{T}\circledast\mathbf{I}_{N_{t,i}}\right)
\mathrm{diag}\left(\mathrm{vec}\left(\widehat{\mathbf{F}}_{\mathrm{RF},i}^{\left(t\right)}\right)\right)\mathrm{vec}\left(j\mathbf{F}_{\mathrm{RF},i}^{\left(t\right)}\right)
\right\|_{\mathrm{F}}^{2}.
\end{split}
\end{equation}

Problem~\eqref{Cachenable29} is convex w.r.t. each individual optimization variable~\cite{BookBoyd2004}. Now, an alternating optimization algorithm can be designed to update the analog precoders and the resulting procedure is summarized in Algorithm~\ref{CachenableAlg02}, where $\eta$ is a contraction factor which compresses the search space. Once  in Step~\ref{CachenableAlg0205} the solution of problem~\eqref{Cachenable29} is obtained,  in Step~\ref{CachenableAlg0206}, $\mathbf{F}_{\mathrm{RF}}^{\left(t+1\right)}$ can be updated using~\eqref{Cachenable25} and $\mathbf{\delta}^{\left(t\right)}$. 
\begin{algorithm}[t]
\caption{Solution of problem~\eqref{Cachenable24}}\label{CachenableAlg02}
\begin{algorithmic}[1]
\STATE Given the digital digital precoding matrices $\mathcal{G}$, $\varepsilon^{\left(0\right)}=0.1$, and $\eta=0.1$.\label{CachenableAlg0201}
\STATE  Set $t=0$ and initialize $\mathbf{F}_{\mathrm{RF}}^{\left(t\right)}$ such that constraints \eqref{Cachenable12b} to \eqref{Cachenable12g} are satisfied.\label{CachenableAlg0202}
\STATE Compute the MMSE-receive filters $\mathcal{X}^{\left(t+1\right)}$ for the given $\mathcal{G}$ and $\mathbf{F}_{\mathrm{RF}}^{\left(t\right)}$ using~\eqref{Cachenable16}.\label{CachenableAlg0203}
\STATE Compute the weight matrices $\mathcal{V}^{\left(t+1\right)}$ for the given $\mathcal{G}$ and $\mathbf{F}_{\mathrm{RF}}^{\left(t\right)}$ using~\eqref{Cachenable19}.\label{CachenableAlg0204}
\STATE Solve problem~\eqref{Cachenable29} to obtain $\mathcal{\delta}^{\left(t\right)}$ and $\mathcal{R}^{\left(t+1\right)}$ for the given $\mathcal{X}^{\left(t+1\right)}$ and  $\mathcal{V}^{\left(t+1\right)}$.\label{CachenableAlg0205}
\STATE Calculate $\mathbf{F}_{\mathrm{RF}}^{\left(t+1\right)}$ with~\eqref{Cachenable25}, $\mathcal{\delta}^{\left(t\right)}$, and $\mathbf{F}_{\mathrm{RF}}^{\left(t\right)}$.\label{CachenableAlg0206}
\STATE If $\exists i\in\mathcal{K}_{\mathrm{R}}$, such that $P_{i}\leqslant p_{i}\left(\mathbf{F}_{\mathrm{RF},i}^{\left(t+1\right)},\mathcal{G}\right)$, then let $\varepsilon^{\left(t\right)}=\eta\varepsilon^{\left(t\right)}$ and go to Step~\ref{CachenableAlg0205}. \label{CachenableAlg0207}
\STATE If $\left| R_{\mathrm{min}}^{\left(t+1\right)}-R_{\mathrm{min}}^{\left(t\right)}\right|\leqslant\epsilon$, then stop the iteration and output $\mathbf{F}_{\mathrm{RF}}^{\left(t+1\right)}$. Otherwise, set $t\leftarrow t+1$, $\varepsilon^{\left(t\right)}=\varepsilon^{\left(t-1\right)}$, and go to Step~\ref{CachenableAlg0203}.\label{CachenableAlg0208}
\end{algorithmic}
\end{algorithm}

In Algorithm~\ref{CachenableAlg02}, the update of the MMSE-receive filters $\mathcal{X}$ and the weight matrices $\mathcal{V}$ does not change the objective value, but may increase the feasible set for improving the minimum user rate. In addition, in problem~\eqref{Cachenable29}, each entry of the analog precoding matrices is updated via a Taylor series approximation in the vicinity of the analog precoder obtained in the previous iteration. This approximation may cause the analog precoding super-matrix $\mathbf{F}_{\mathrm{RF}}$ obtained in Step~\ref{CachenableAlg0206} of Algorithm~\ref{CachenableAlg02} to not satisfy the power constraint in~\eqref{Cachenable12f}.  If the power constraint of a certain eRRH is violated, the approximation vicinity of the analog precoder obtained in the previous iteration has to be adjusted, i.e., the search space has to be compressed.  To this end, an additional space compression step, i.e., Step~\ref{CachenableAlg0207} of Algorithm~\ref{CachenableAlg02}, is introduced to adjust the search radius until the power constraint is satisfied.  In the worst case, the analog precoding super-matrix $\mathbf{F}_{\mathrm{RF}}$ remains unchanged from one iteration to the next, i.e., the objective value of problem~\eqref{Cachenable29} s the unchanged. Therefore, Step~\ref{CachenableAlg0206} of Algorithm~\ref{CachenableAlg02} generates a non-decreasing sequence of the objective value of problem~\eqref{Cachenable29}. As a consequence, Algorithm~2 yields the following non-decreasing sequence of objective values
$$R_{\mathrm{min}}^{\left(0\right)}\leq R_{\mathrm{min}}^{\left(1\right)}\leq\cdots\leq R_{\mathrm{min}}^{\left(t\right)}\leq R_{\mathrm{min}}^{\left(t+1\right)}\leq\cdots$$
Combining this with the bounded objective function of problem~\eqref{Cachenable29}, the convergence of Algorithm~\ref{CachenableAlg02} to a fixed point is guaranteed~\cite{Bibby1974}. Given a tolerance $\epsilon$, Algorithm~\ref{CachenableAlg02} has converged when the error $\left| R_{\mathrm{min}}^{\left(t+1\right)}-R_{\mathrm{min}}^{\left(t\right)}\right|$ falls below $\epsilon$.

Similar to Algorithm~\ref{CachenableAlg01}, Algorithm~\ref{CachenableAlg02} also requires matrix multiplications, matrix inversions, and solving a convex optimzation problem.  Steps~\ref{CachenableAlg0203} and~\ref{CachenableAlg0204} of Algorithm~\ref{CachenableAlg02} involve the same number of operations as Steps~\ref{CachenableAlg0103} and~\ref{CachenableAlg0104} of Algorithm~\ref{CachenableAlg01}. The complexity of solving the convex optimization problem is $\mathit{O}\left(\phi_{2}\right)$, where $\phi_{2}=\left(LK_{\mathrm{U}}+N_{\mathrm{tm}}N_{\mathrm{tRF}}K_{\mathrm{R}}\right)^{3.5}$ and $N_{\mathrm{tm}}=\max\limits_{i\in\mathcal{K}_{\mathrm{R}}}N_{\mathrm{t},i}$~\cite{MathPotra2000}. Hence, the overall computational complexity of Algorithm~\ref{CachenableAlg02} is $\mathit{O}\left(\kappa_{2}\left(K_{\mathrm{U}}LN_{v,1}+\phi_{2}\right)\right)$, where $\kappa_{2}$ denotes the total number of iterations of Algorithm~\ref{CachenableAlg02}.

\subsection*{D. Optimization of the Hybrid Precoder}

In the previous two subsections, alternating optimization algorithms have been proposed to optimize the digital/analog precoding matrices for given analog/digital precoders. Thus, the joint design of the digital precoder $\mathcal{G}$ and the analog precoding super-matrix $\mathbf{F}_{\mathrm{RF}}$ can be implemented by running Algorithm~\ref{CachenableAlg01} and Algorithm~\ref{CachenableAlg02} in an alternating manner. The convergence proofs of both algorithms show that each algorithm generates a non-decreasing sequence. In the alternating execution of Algorithm~\ref{CachenableAlg01} and Algorithm~\ref{CachenableAlg02}, the output of one algorithm serves as input for the other algorithm. Therefore, the alternating optimization of digital precoder $\mathcal{G}$ and analog precoding super-matrix $\mathbf{F}_{\mathrm{RF}}$ produces a non-decreasing sequence $R_{\mathrm{min}}$. Since the objective function of problem~\eqref{Cachenable12} is a bounded function due to the limited transmit power, the objective values generated by the alternating execution of Algorithm~\ref{CachenableAlg01} and Algorithm~\ref{CachenableAlg02} form a convergent sequence\footnote{Note that the proving the optimality, such as local/global optimality or satisfying Karush-Kuhn-Tucker conditions, of the solution obtained by the alternating execution of Algorithm~\ref{CachenableAlg01} and Algorithm~\ref{CachenableAlg02} is a challenging task and is left for future work.}~\cite{Bibby1974}.

\section*{\sc \uppercase\expandafter{\romannumeral4}. Hybrid Precoding Design for Soft Fronthaul Information Transfer}

In this section, we consider the design of hybrid precoding for CeMm-RANs with SFIT, where the BBU transfers a quantized version of the precoded signals of the missing files to the eRRHs.  Accordingly, signal $\mathbf{x}_{i}$ transmitted by eRRH $i$ in the downlink is the superposition of two signals, where one signal is locally precoded based on the content of the cache, whereas the other signal is precoded at the BBU and quantized for transmission over the fronthaul link~\cite{TWCPark2016}, we have
\begin{equation}\label{Cachenable32}
\mathbf{x}_{i}=\mathbf{F}_{\mathrm{RF},i}\left(\sum\limits_{f\in \mathcal{F}_{\mathrm{req}}}\sum\limits_{l\in \mathcal{L}}c_{f,l,i}\mathbf{G}_{f,l,i}\mathbf{s}_{f,l}+\widehat{\mathbf{x}}_{i}\right).
\end{equation}
Here, $\widehat{\mathbf{x}}_{i}=\widetilde{\mathbf{x}}_{i}+\mathbf{z}_{i}$ is the quantized signal received from the BBU via the fronthaul link,
where the quantization noise $\mathbf{z}_{i}\in\mathds{C}^{N_{\mathrm{tRF},i}\times 1}$ is assumed to be independent of $\widetilde{\mathbf{x}}_{i}$ and distributed as $\mathbf{z}_{i}\sim\mathcal{CN}\left(\mathbf{0},\mathbf{\Omega}_{i}\right)$. We further assume that the quantization noise $\mathbf{z}_{i}$ is independent across the eRRHs, i.e., the signals intended for different eRRHs are quantized independently~\cite{TVTKang2016}. $\widetilde{\mathbf{x}}_{i}$ is the precoded signal that includes the requested subfiles which are not stored at eRRH $i$ and is given by
\begin{equation}\label{Cachenable34}
\widetilde{\mathbf{x}}_{i}=\sum\limits_{f\in \mathcal{F}_{\mathrm{req}}}\sum\limits_{l\in \mathcal{L}}\overline{c}_{f,l,i}\mathbf{U}_{f,l,i}\mathbf{s}_{f,l},
\end{equation}
where  $\mathbf{U}_{f,l,i}\in\mathbb{C}^{N_{\mathrm{tRF},i}\times d_{f,l}}$ is the digital precoding matrix at the BBU for baseband signal $\mathbf{s}_{f,l}$ representing subfile $\left(f,l\right)$ which is not available at eRRH $i$. The rate on the fronthaul link of eRRH $i$ can therefore be expressed as
\begin{equation}\label{Cachenable35}
g_{i}\left(\mathcal{U},\mathcal{O}\right)\triangleq I\left(\widetilde{\mathbf{x}}_{i}; \widehat{\mathbf{x}}_{i}\right)
=\log\det\left(\sum\limits_{f\in \mathcal{F}_{\mathrm{req}}}\sum\limits_{l\in \mathcal{L}}\overline{c}_{f,l,i}\mathbf{U}_{f,l,i}\mathbf{U}_{f,l,i}^{H}+\mathbf{\Omega}_{i}\right)
-\log\det\left(\mathbf{\Omega}_{i}\right),
\end{equation}
where $\mathcal{U}\triangleq\left\{\mathbf{U}_{f,l,i}\right\}_{f\in \mathcal{F}_{\mathrm{req}},l\in \mathcal{L},i\in\mathcal{K}_{\mathrm{R}}}$ and $\mathcal{O}\triangleq\left\{\mathbf{\Omega}_{i}\right\}_{i\in\mathcal{K}_{\mathrm{R}}}$. According to~\cite[Ch.~3]{BookGamal2011}, to reliably recover signal $\widehat{\mathbf{x}}_{i}$ at eRRH $i$, the constraint $g_{i}\left(\mathcal{U},\mathcal{O}\right)\leqslant \log\left(2\right)C_{i}$ has to be satisfied.

For SFIT, $\overline{\mathbf{U}}_{f,l,i}$ is defined as $\overline{\mathbf{U}}_{f,l,i}=c_{f,l,i}\mathbf{G}_{f,l,i}+\overline{c}_{f,l,i}\mathbf{U}_{f,l,i}$, $f\in \mathcal{F}_{\mathrm{req}}$, $l\in \mathcal{L}$, $\forall i\in\mathcal{K}_{\mathrm{R}}$. Exploiting permutation matrix $\mathbf{P}_{i}$ and $\overline{\mathbf{U}}_{f,l}$, $g_{i}\left(\mathcal{U}, \mathcal{O}\right)$ can be replaced by
\begin{equation}\label{Cachenable36}
\overline{g}_{i}\left(\mathcal{U},\mathcal{G}, \mathcal{O}\right)\triangleq\log\det\left(\sum\limits_{f\in \mathcal{F}_{\mathrm{req}}}\sum\limits_{l\in \mathcal{L}}\overline{c}_{f,l,i}\mathbf{P}_{i}^{H}\overline{\mathbf{U}}_{f,l}\overline{\mathbf{U}}_{f,l}^{H}\mathbf{P}_{i}+\mathbf{\Omega}_{i}\right)
-\log\det\left(\mathbf{\Omega}_{i}\right),
\end{equation}
since if $\overline{c}_{f,l,i}=1$, then $\mathbf{P}_{i}^{H}\overline{\mathbf{U}}_{f,l}=\overline{\mathbf{U}}_{f,l,i}=\mathbf{U}_{f,l,i}$. The received signal $\mathbf{y}_{k}$ at user $k$ for requested file $f_{k}$ can be expressed as
\begin{equation}\label{Cachenable37}
\mathbf{y}_{k}=\sum\limits_{l\in \mathcal{L}}\overline{\mathbf{H}}_{k,f_{k},l}\mathbf{s}_{f_{k},l}+
\sum\limits_{f\in\mathcal{F}_{\mathrm{req}}\setminus\left\{f_{k}\right\}}\sum\limits_{l\in \mathcal{L}}\overline{\mathbf{H}}_{k,f,l}\mathbf{s}_{f,l}
+\mathbf{H}_{k}\mathbf{F}_{\mathrm{RF}}\mathbf{z}+\mathbf{n}_{k},
\end{equation}
where $\mathbf{z}=\left[\mathbf{z}_{1}^{H},\cdots,\mathbf{z}_{K_{\mathrm{R}}}^{H}\right]^{H}$. Therefore, the achievable data rate for subfile $\left(f_{k},l\right)$ at user $k$ is given by
\begin{equation}\label{Cachenable38}
\widehat{q}_{k,l}\left(\mathbf{F}_{\mathrm{RF}},\mathcal{U},\mathcal{G},\mathcal{O}\right)\triangleq
\log\det\left(\mathbf{I}_{N_{\mathrm{r},k}\times N_{\mathrm{r},k}}+\overline{\mathbf{H}}_{k,f_{k},l}\overline{\mathbf{H}}_{k,f_{k},l}^{H}\mathbf{\Pi}_{k,l}^{-1}\right).
\end{equation}
Different from~\eqref{Cachenable11}, for SFIT, $\mathbf{\Pi}_{k,l}$ in~\eqref{Cachenable38} is given by
\begin{equation}\label{Cachenable39}
\begin{split}
\mathbf{\Pi}_{k,l}&=\sum\limits_{m\in \mathcal{L}\setminus\left\{l\right\}}\overline{\mathbf{H}}_{k,f_{k},m}\overline{\mathbf{H}}_{k,f_{k},m}^{H}
+\sum\limits_{f\in\mathcal{F}_{\mathrm{req}}\setminus\left\{f_{k}\right\}}\sum\limits_{\tau\in \mathcal{L}}\overline{\mathbf{H}}_{k,f,\tau}\overline{\mathbf{H}}_{k,f,\tau}^{H}
+\mathbf{H}_{k}\mathbf{F}_{\mathrm{RF}}\overline{\mathbf{\Omega}}\mathbf{F}_{\mathrm{RF}}^{H}\mathbf{H}_{k}^{H}+\sigma_{k}^{2}\mathbf{I}_{N_{r,k}},
\end{split}
\end{equation}
where $\overline{\mathbf{\Omega}}=\mathrm{diag}\left(\mathbf{\Omega}_{1},\cdots,\mathbf{\Omega}_{K_{\mathrm{R}}}\right)$.

\subsection*{A. Problem Formulation}

The objective is again the maximization of  the minimum user rate $R_{\mathrm{min}}\triangleq\min\limits_{f\in\mathcal{F}_{\mathrm{req}}} R_{f}$ under the fronthaul capacity, eRRH transmit power, and constant-modulus precoder constraints. The  resulting design problem is formulated as
\begin{subequations}\label{Cachenable40}
\begin{align}
&\max_{\mathbf{F}_{\mathrm{RF}},\mathcal{U},\mathcal{G},\mathcal{O}, \mathcal{R}}R_{\mathrm{min}},\label{Cachenable40a}\\
\mathrm{s.t.}~
&\log\left(2\right)R_{f_{k},l}\leq \widehat{q}_{k,l}\left(\mathbf{F}_{\mathrm{RF}},\mathcal{U},\mathcal{G},\mathcal{O}\right), \forall k\in\mathcal{K}_{\mathrm{U}}, l\in\mathcal{L},\label{Cachenable40b}\\
&\overline{g}_{i}\left(\mathcal{U},\mathcal{G},\mathcal{O}\right)\leqslant \log\left(2\right)C_{i}, \mathbf{\Omega}_{i}\succeq 0, \forall i\in\mathcal{K}_{\mathrm{R}},\label{Cachenable40c}\\
&\overline{p}_{i}\left(\mathbf{F}_{\mathrm{RF},i},\mathcal{U},\mathcal{G},\mathbf{\Omega}_{i}\right)\leq P_{i}, \forall i\in\mathcal{K}_{\mathrm{R}},\label{Cachenable40d}\\
&\eqref{Cachenable12c},\eqref{Cachenable12d},\eqref{Cachenable12g},\label{Cachenable40e}
\end{align}
\end{subequations}
where $\overline{p}_{i}\left(\mathbf{F}_{\mathrm{RF},i},\mathcal{U},\mathcal{G},\mathbf{\Omega}_{i}\right)$ is given by
\begin{equation}\label{Cachenable41}
\begin{split}
\overline{p}_{i}\left(\mathbf{F}_{\mathrm{RF},i},\mathcal{U},\mathcal{G},\mathbf{\Omega}_{i}\right)&\triangleq
\sum\limits_{f\in\mathcal{F}_{\mathrm{req}}}\sum\limits_{l\in\mathcal{L}}\mathrm{tr}\left(\mathbf{F}_{\mathrm{RF},i}\overline{\mathbf{U}}_{f,l,i}
\overline{\mathbf{U}}_{f,l,i}^{H}\mathbf{F}_{\mathrm{RF},i}^{H}\right)+\mathrm{tr}\left(\mathbf{F}_{\mathrm{RF},i}
\mathbf{\Omega}_{i}\mathbf{F}_{\mathrm{RF},i}^{H}\right).
\end{split}
\end{equation}
Similar to problem~\eqref{Cachenable12}, problem~\eqref{Cachenable40} is difficult to solve due to the non-convexity of the achievable data rate in~\eqref{Cachenable40b}, the fronthaul capacity constraint in~\eqref{Cachenable40c}, the constant-modulus requirement on the entries of the analog precoders in~\eqref{Cachenable12g}, and the strong coupling between the analog precoding matrices and the digital precoding matrices. To overcome the non-convexity of $\overline{g}_{i}\left(\mathcal{U},\mathcal{G},\mathcal{O}\right)$, exploiting the concavity of $\log\det\left(\cdot\right)$, we have $\overline{g}_{i}\left(\mathcal{U},\mathcal{G},\mathcal{O}\right)\leqslant\widetilde{g}_{i}
\left(\mathcal{U},\mathcal{G},\mathcal{O},\mathcal{Z}\right)$~\cite{TSPZhou2016},
where $\widetilde{g}_{i}\left(\mathcal{U},\mathcal{G},\mathcal{O},\mathcal{Z}\right)$ is given by:
\begin{equation}\label{Cachenable42}
\begin{split}
\widetilde{g}_{i}\left(\mathcal{U},\mathcal{G},\mathcal{O},\mathcal{Z}\right)\triangleq &\log\det\left(\mathbf{\Sigma}_{i}\right)
-N_{\mathrm{tRF},i}-\log\det\left(\mathbf{\Omega}_{i}\right)\\
&+\mathrm{tr}\left(\mathbf{\Sigma}_{i}^{-1}\left(\sum\limits_{f\in \mathcal{F}_{\mathrm{req}}}\sum\limits_{l\in \mathcal{L}}\overline{c}_{f,l,i}\mathbf{P}_{i}^{H}\overline{\mathbf{U}}_{f,l}\overline{\mathbf{U}}_{f,l}^{H}\mathbf{P}_{i}+\mathbf{\Omega}_{i}\right)\right)
\end{split}
\end{equation}
with auxiliary variable $\mathcal{Z}=\left\{\mathbf{\Sigma}_{i}\succ\mathbf{0}\right\}_{i\in\mathcal{K}_{\mathrm{R}}}$. Constraint~\eqref{Cachenable40c} and $\widetilde{g}_{i}\left(\mathcal{U},\mathcal{G},\mathcal{O},\mathcal{Z}\right)\leqslant\log\left(2\right) C_{i}$ are equivalent when
\begin{equation}\label{Cachenable44}
\mathbf{\Sigma}_{i}=\sum\limits_{f\in \mathcal{F}_{\mathrm{req}}}\sum\limits_{l\in \mathcal{L}}\overline{c}_{f,l,i}\mathbf{P}_{i}^{H}\overline{\mathbf{U}}_{f,l}\overline{\mathbf{U}}_{f,l}^{H}\mathbf{P}_{i}+\mathbf{\Omega}_{i}
=\sum\limits_{f\in \mathcal{F}_{\mathrm{req}}}\sum\limits_{l\in \mathcal{L}}\overline{c}_{f,l,i}\mathbf{U}_{f,l,i}\mathbf{U}_{f,l,i}^{H}+\mathbf{\Omega}_{i}.
\end{equation}
Note that although $\widetilde{g}_{i}\left(\mathcal{U},\mathcal{G},\mathcal{O},\mathcal{Z}\right)$ is not jointly convex w.r.t. variables $\mathcal{U}$, $\mathcal{G}$, $\mathcal{O}$, and $\mathcal{Z}$, it is jointly convex w.r.t. $\mathcal{U}$, $\mathcal{G}$, and $\mathcal{O}$ for a given $\mathcal{Z}$. In the sequel, we replace constraint~\eqref{Cachenable40c} with $\widetilde{g}_{i}\left(\mathcal{U},\mathcal{G},\mathcal{O},\mathcal{Z}\right)\leqslant \log\left(2\right) C_{i}$.

To overcome the non-convexity of the achievable data rate $\widehat{q}_{k,l}\left(\mathbf{F}_{\mathrm{RF}},\mathcal{U},\mathcal{G},\mathcal{O}\right)$ for subfile $\left(f_{k},l\right)$, we approximate $\widehat{q}_{k,l}\left(\mathbf{F}_{\mathrm{RF}},\mathcal{U},\mathcal{G},\mathcal{O}\right)$ by function $\breve{q}_{k,l}\left(\mathbf{F}_{\mathrm{RF}},\mathcal{U},\mathcal{G},\mathcal{O}, \mathcal{V}, \mathcal{X}\right)$, which is defined as
\begin{equation}\label{Cachenable47}
\breve{q}_{k,l}\left(\mathbf{F}_{\mathrm{RF}},\mathcal{U},\mathcal{G},\mathcal{O}, \mathcal{V}, \mathcal{X}\right)\triangleq
\log\det\left(\mathbf{\Upsilon}_{k,l}\right)-\mathrm{tr}\left(\mathbf{\Upsilon}_{k,l}\mathbf{E}_{k,l}\right)
+d_{f_{k},l}.
\end{equation}
The MSE matrix $\mathbf{E}_{k,l}$ for recovering subfile $\left(f_{k},l\right)$ and the MMSE filter $\mathbf{\Xi}_{k,l}^{\mathrm{MMSE}}$ at user $k$ are calculated as~\eqref{Cachenable14} and~\eqref{Cachenable16}, respectively, where $\mathbf{\Lambda}_{k,l}$ is here given by
\begin{equation}\label{Cachenable45}
\mathbf{\Lambda}_{k,l}=\sum\limits_{f\in\mathcal{F}_{\mathrm{req}}}\sum\limits_{\tau\in \mathcal{L}}\overline{\mathbf{H}}_{k,f,\tau}\overline{\mathbf{H}}_{k,f,\tau}^{H}
+\mathbf{H}_{k}\mathbf{F}_{\mathrm{RF}}\overline{\mathbf{\Omega}}\mathbf{F}_{\mathrm{RF}}^{H}\mathbf{H}_{k}^{H}+\sigma_{k}^{2}\mathbf{I}_{N_{r,k}}.
\end{equation}
Thus, optimization problem~\eqref{Cachenable40} can be reformulated as follows:
\begin{subequations}\label{Cachenable46}
\begin{align}
&\max_{\mathbf{F}_{\mathrm{RF}},\mathcal{U},\mathcal{G},\mathcal{O}, \mathcal{R}, \mathcal{V},\mathcal{X},\mathcal{Z}}R_{\mathrm{min}},\label{Cachenable46a}\\
\mathrm{s.t.}~
&\log\left(2\right)R_{f_{k},l}\leq \breve{q}_{k,l}\left(\mathbf{F}_{\mathrm{RF}},\mathcal{U},\mathcal{G},\mathcal{O}, \mathcal{V}, \mathcal{X}\right), \forall k\in\mathcal{K}_{\mathrm{U}}, l\in\mathcal{L},\label{Cachenable46b}\\
&\widetilde{g}_{i}\left(\mathcal{U},\mathcal{G},\mathcal{O},\mathcal{Z}\right)\leqslant \log\left(2\right)C_{i}, \mathbf{\Omega}_{i}\succeq 0, \forall i\in\mathcal{K}_{\mathrm{R}},\label{Cachenable46c}\\
&\eqref{Cachenable12c},\eqref{Cachenable12d},\eqref{Cachenable12g},\eqref{Cachenable40d}.\label{Cachenable46d}
\end{align}
\end{subequations}
In the following, we adopt alternating optimization to solve problem~\eqref{Cachenable46}.
\subsection*{B. Optimization of Digital Precoder and Quantization Noise Covariance Matrix}
Similar to problem~\eqref{Cachenable12}, problem~\eqref{Cachenable46} can be solved by using alternating optimization methods.  For given analog precoders, the digital precoders are optimized by solving the following problem:
\begin{equation}\label{Cachenable48}
\max_{\mathcal{U},\mathcal{G},\mathcal{O}, \mathcal{R}, \mathcal{V},\mathcal{X},\mathcal{Z}}R_{\mathrm{min}}~
\mathrm{s.t.}~\eqref{Cachenable12c},\eqref{Cachenable12d},\eqref{Cachenable40d},\eqref{Cachenable46b},\eqref{Cachenable46c}.
\end{equation}
The detailed steps used for solving problem~\eqref{Cachenable48} are summarized in Algorithm~\ref{CachenableAlg03}. Similar to Algorithm~\ref{CachenableAlg01}, Algorithm~\ref{CachenableAlg03} yields a non-decreasing sequence of objective values for problem~\eqref{Cachenable48}. Recalling the bounded objective function of problem~\eqref{Cachenable48}, Algorithm~\ref{CachenableAlg03} is guaranteed to converge~\cite{Bibby1974}. Following the same arguments as those in~\cite{OperalMarks1977}, we can prove that Algorithm~\ref{CachenableAlg03} converges to a stationary point of optimization problem~\eqref{Cachenable48}. The computational complexity of Algorithm~\ref{CachenableAlg03} can be analyzed in a similar manner as that of Algorithm~\ref{CachenableAlg01}.
\begin{algorithm}[t]
\caption{Solution of problem~\eqref{Cachenable48}}\label{CachenableAlg03}
\begin{algorithmic}[1]
\STATE Analog precoding super-matrix $\mathbf{F}_{\mathrm{RF}}$ is given.\label{CachenableAlg0301}
\STATE  Set $t=0$ and initialize $\mathcal{U}^{\left(t\right)}$, $\mathcal{G}^{\left(t\right)}$, $\mathcal{O}^{\left(t\right)}$, and $\mathcal{R}^{\left(t\right)}$ such that \eqref{Cachenable46b} to \eqref{Cachenable46d} are satisfied.\label{CachenableAlg0302}
\STATE Calculate $\mathcal{Z}^{\left(t\right)}$ for the given $\mathcal{U}^{\left(t\right)}$ and $\mathcal{O}^{\left(t\right)}$ using~\eqref{Cachenable44}.\label{CachenableAlg0306}
\STATE Compute  the MMSE filters $\mathcal{X}^{\left(t+1\right)}$ for the given $\mathcal{U}^{\left(t\right)}$, $\mathcal{G}^{\left(t\right)}$, and $\mathcal{O}^{\left(t\right)}$  using \eqref{Cachenable16} and~\eqref{Cachenable45}.\label{CachenableAlg0303}
\STATE Compute the weight matrices $\mathcal{V}^{\left(t+1\right)}$ for the given $\mathcal{U}^{\left(t\right)}$, $\mathcal{G}^{\left(t\right)}$, and $\mathcal{O}^{\left(t\right)}$ using \eqref{Cachenable19} and~\eqref{Cachenable45}.\label{CachenableAlg0304}
\STATE Solve problem~\eqref{Cachenable48} to obtain $\mathcal{U}^{\left(t+1\right)}$, $\mathcal{G}^{\left(t+1\right)}$, $\mathcal{O}^{\left(t+1\right)}$, and $\mathcal{R}^{\left(t+1\right)}$ for the given $\mathcal{Z}^{\left(t\right)}$, $\mathcal{X}^{\left(t+1\right)}$, and  $\mathcal{V}^{\left(t+1\right)}$.\label{CachenableAlg0305}
\STATE If $\left| R_{\mathrm{min}}^{\left(t+1\right)}-R_{\mathrm{min}}^{\left(t\right)}\right|\leqslant\epsilon$, then stop the iteration and output $\mathcal{U}^{\left(t+1\right)}$, $\mathcal{G}^{\left(t+1\right)}$, $\mathcal{O}^{\left(t+1\right)}$, and $\mathcal{R}^{\left(t+1\right)}$. Otherwise, set $t\leftarrow t+1$ and go to Step~\ref{CachenableAlg0306}.\label{CachenableAlg0307}
\end{algorithmic}
\end{algorithm}
\subsection*{C. Optimization of Analog Precoder}

Different from HFIT, by compressing the precoded subfiles that are not stored at the eRRHs, quantization noise is introduced. Therefore, we need to derive the expression of $\mathrm{tr}\left(\mathbf{\Upsilon}_{k,l}^{\left(t\right)}\mathbf{E}_{k,l}^{\left(t+1\right)}\right)$ for SFIT. With $\mathbf{\Omega}_{i}=\widetilde{\mathbf{\Omega}}_{i}\widetilde{\mathbf{\Omega}}_{i}^{H}$ and  $\widetilde{\mathbf{\Omega}}=\mathrm{diag}\left(\widetilde{\mathbf{\Omega}}_{1},\cdots,\widetilde{\mathbf{\Omega}}_{i},\cdots,\widetilde{\mathbf{\Omega}}_{K_{\mathrm{R}}}\right)$, $\mathrm{tr}\left(\mathbf{\Upsilon}_{k,l}^{\left(t\right)}\mathbf{E}_{k,l}^{\left(t+1\right)}\right)$ can be expressed as follows:
\begin{equation}\label{Cachenable50}
\begin{split}
&\mathrm{tr}\left(\mathbf{\Upsilon}_{k,l}^{\left(t\right)}\mathbf{E}_{k,l}^{\left(t+1\right)}\right)
=\left\|\mathbf{b}_{k,l,f_{k},l}^{\left(t\right)}-\widetilde{\mathbf{c}}_{k,l,f_{k},l}^{\left(t\right)}\right\|_{\mathrm{F}}^{2}
+\sum\limits_{m\in\mathcal{L}\setminus\left\{l\right\}}\left\|\mathbf{c}_{k,l,f_{k},m}^{\left(t\right)}
+\widetilde{\mathbf{c}}_{k,l,f_{k},m}^{\left(t\right)}\right\|_{\mathrm{F}}^{2}\\
&+\sum\limits_{f\in\mathcal{F}_{\mathrm{req}}\setminus\left\{f_{k}\right\}}\sum\limits_{\tau\in \mathcal{L}}\left\|\mathbf{c}_{k,l,f,\tau}^{\left(t\right)}+\widetilde{\mathbf{c}}_{k,l,f,\tau}^{\left(t\right)}\right\|_{\mathrm{F}}^{2}+\left\|\mathbf{d}_{k,l}^{\left(t\right)}+\widetilde{\mathbf{d}}_{k,l}^{\left(t\right)}\right\|_{\mathrm{F}}^{2}
+\mathrm{tr}\left(\sigma_{k}^{2}\left(\mathbf{\Xi}_{k,l}^{\left(t\right)}\right)^{H}\mathbf{\Xi}_{k,l}^{\left(t\right)}\mathbf{\Upsilon}_{k,l}^{\left(t\right)}\right),
\end{split}
\end{equation}
where $\mathbf{D}_{k,l}^{\left(t\right)}=\left(\left(\overline{\mathbf{\Upsilon}}_{k,l}^{\left(t\right)}\right)^{T}\left(\mathbf{\Xi}_{k,l}^{\left(t\right)}\right)^{T}\mathbf{H}_{k}^{*}\right)\circledast\widetilde{\mathbf{\Omega}}^{H}$, $\mathbf{d}_{k,l}^{\left(t\right)}=\mathbf{D}_{k,l}^{\left(t\right)}\mathrm{vec}\left(\left(\mathbf{F}_{\mathrm{RF}}^{\left(t\right)}\right)^{H}\right)$, and
$$\widetilde{\mathbf{d}}_{k,l}^{\left(t\right)}=\mathbf{D}_{k,l}^{\left(t\right)}\mathrm{vec}\left(\left(\widetilde{\mathbf{F}}_{\mathrm{RF}}^{\left(t\right)}\right)^{H}\right)=\mathbf{D}_{k,l}^{\left(t\right)}\mathrm{diag}\left(\mathrm{vec}\left(\left(\widehat{\mathbf{F}}_{\mathrm{RF}}^{\left(t\right)}\right)^{H}\right)\right)\mathrm{vec}\left(\left(j\mathbf{F}_{\mathrm{RF}}^{\left(t\right)}\right)^{H}\right).$$
For SFIT, $\widehat{p}_{i}^{\left(t\right)}\left(\mathcal{\delta}^{\left(t\right)}\right)$ is redefined as follows:
\begin{equation}\label{Cachenable51}
\begin{split}
&\widehat{p}_{i}^{\left(t\right)}\left(\mathcal{\delta}^{\left(t\right)}\right)\triangleq\sum\limits_{f\in\mathcal{F}_{\mathrm{req}}}
\sum\limits_{l\in\mathcal{L}}\left\|\left(\mathbf{F}_{\mathrm{RF},i}^{\left(t\right)}+\widehat{\mathbf{F}}_{\mathrm{RF},i}^{\left(t\right)}\circ j\mathbf{F}_{\mathrm{RF},i}^{\left(t\right)}\right)\overline{\mathbf{U}}_{f,l,i}\right\|_{\mathrm{F}}^{2}+\left\|\left(\mathbf{F}_{\mathrm{RF},i}^{\left(t\right)}+\widehat{\mathbf{F}}_{\mathrm{RF},i}^{\left(t\right)}\circ j\mathbf{F}_{\mathrm{RF},i}^{\left(t\right)}\right)\widetilde{\mathbf{\Omega}}_{i}\right\|_{\mathrm{F}}^{2}\\
&=\sum\limits_{f\in\mathcal{F}_{\mathrm{req}}}
\sum\limits_{l\in\mathcal{L}}\left\|\mathrm{vec}\left(\mathbf{F}_{\mathrm{RF},i}^{\left(t\right)}\overline{\mathbf{U}}_{f,l,i}\right)+
\left(\overline{\mathbf{U}}_{f,l,i}^{T}\circledast\mathbf{I}_{N_{t,i}}\right)
\mathrm{diag}\left(\mathrm{vec}\left(\widehat{\mathbf{F}}_{\mathrm{RF},i}^{\left(t\right)}\right)\right)\mathrm{vec}\left(j\mathbf{F}_{\mathrm{RF},i}^{\left(t\right)}\right)
\right\|_{\mathrm{F}}^{2}\\
&+\left\|\mathrm{vec}\left(\mathbf{F}_{\mathrm{RF},i}^{\left(t\right)}\widetilde{\mathbf{\Omega}}_{i}\right)+
\left(\widetilde{\mathbf{\Omega}}_{i}^{T}\circledast\mathbf{I}_{N_{t,i}}\right)
\mathrm{diag}\left(\mathrm{vec}\left(\widehat{\mathbf{F}}_{\mathrm{RF},i}^{\left(t\right)}\right)\right)\mathrm{vec}\left(j\mathbf{F}_{\mathrm{RF},i}^{\left(t\right)}\right)
\right\|_{\mathrm{F}}^{2}.
\end{split}
\end{equation}
The analog precoding matrices can again be obtained by using an iterative optimization method and in the $t$-th iteration the following problem is solved:
\begin{subequations}\label{Cachenable52}
\begin{align}
&\max_{\mathcal{\delta}^{\left(t\right)},\mathcal{R}, \mathcal{V},\mathcal{X}}R_{\mathrm{min}},\label{Cachenable52a}\\
\mathrm{s.t.}~&\log\left(2\right)R_{f_{k},l}\leq \widehat{q}_{k,l}^{\left(t\right)}\left(\mathcal{\delta}^{\left(t\right)}, \mathcal{V}, \mathcal{X}\right), \forall k\in\mathcal{K}_{\mathrm{U}}, l\in\mathcal{L},\label{Cachenable52b}\\
&\eqref{Cachenable12c},\eqref{Cachenable12d}, \widehat{p}_{i}^{\left(t\right)}\left(\mathcal{\delta}^{\left(t\right)}\right)\leq P_{i}, \forall i\in\mathcal{K}_{\mathrm{R}},\label{Cachenable52c}\\
&\left|\delta_{m_{i},n_{i}}^{\left(t\right)}\right|\leqslant\varepsilon^{\left(t\right)},\forall i\in\mathcal{K}_{\mathrm{R}}, m_{i}\in\mathcal{N}_{\mathrm{R}, i}, n_{i}\in\mathcal{N}_{\mathrm{C}, i},\label{Cachenable52d}
\end{align}
\end{subequations}
with $\mathrm{tr}\left(\mathbf{\Upsilon}_{k,l}^{\left(t\right)}\mathbf{E}_{k,l}^{\left(t+1\right)}\right)$ and $\widehat{p}_{i}^{\left(t\right)}\left(\mathcal{\delta}^{\left(t\right)}\right)$ given in~\eqref{Cachenable50} and~\eqref{Cachenable51}, respectively. The steps used to solve problem~\eqref{Cachenable52} are summarized in Algorithm~\ref{CachenableAlg04}. The analysis of the convergence and the computational complexity of Algorithm~\ref{CachenableAlg04} are similar to those of Algorithm~\ref{CachenableAlg02}.
\begin{algorithm}[ht]
\caption{Solution of problem~\eqref{Cachenable52}}\label{CachenableAlg04}
\begin{algorithmic}[1]
\STATE Digital precoding matrices $\mathcal{U}$, $\mathcal{G}$ and the noise covariance matrices $\mathcal{O}$ are given. \label{CachenableAlg0401}
\STATE  Set $t=0$, $\varepsilon^{\left(t\right)}=0.1$ and $\eta=0.1$. Initialize $\mathbf{F}_{\mathrm{RF}}^{\left(t\right)}$ such that \eqref{Cachenable46b} to \eqref{Cachenable46d} are satisfied.\label{CachenableAlg0402}
\STATE Compute the MMSE filters $\mathcal{X}^{\left(t+1\right)}$ for the given $\mathcal{U}$, $\mathcal{G}$, $\mathcal{O}$, and $\mathbf{F}_{\mathrm{RF}}^{\left(t\right)}$ using \eqref{Cachenable16} and~\eqref{Cachenable45}.\label{CachenableAlg0403}
\STATE Calculate the weight matrices $\mathcal{V}^{\left(t+1\right)}$ for the given $\mathcal{U}$, $\mathcal{G}$, $\mathcal{O}$, and $\mathbf{F}_{\mathrm{RF}}^{\left(t\right)}$ using \eqref{Cachenable19} and~\eqref{Cachenable45}.\label{CachenableAlg0404}
\STATE Solve problem~\eqref{Cachenable52} to obtain $\mathcal{\delta}^{\left(t\right)}$ and $\mathcal{R}^{\left(t+1\right)}$ for the given $\mathcal{U}$, $\mathcal{G}$, $\mathcal{O}$, $\mathcal{X}^{\left(t+1\right)}$, and  $\mathcal{V}^{\left(t+1\right)}$.\label{CachenableAlg0405}
\STATE Calculate $\mathbf{F}_{\mathrm{RF}}^{\left(t+1\right)}$ according to~\eqref{Cachenable25} with $\mathcal{\delta}^{\left(t\right)}$ and $\mathbf{F}_{\mathrm{RF}}^{\left(t\right)}$.\label{CachenableAlg0406}
\STATE If $\exists~i\in\mathcal{K}_{\mathrm{R}}$, such that $P_{i}\leqslant \overline{p}_{i}\left(\mathbf{F}_{\mathrm{RF},i}^{\left(t+1\right)},\mathcal{U},\mathcal{G},\mathbf{\Omega}_{i}\right)$, then let $\varepsilon^{\left(t\right)}=\eta\varepsilon^{\left(t\right)}$ and go Step~\ref{CachenableAlg0405}.\label{CachenableAlg0407}
\STATE If $\left| R_{\mathrm{min}}^{\left(t+1\right)}-R_{\mathrm{min}}^{\left(t\right)}\right|\leqslant\epsilon$, then stop the iteration and output $\mathbf{F}_{\mathrm{RF}}^{\left(t+1\right)}$. Otherwise, set $t\leftarrow t+1$, $\varepsilon^{\left(t\right)}=\varepsilon^{\left(t-1\right)}$, and go to Step~\ref{CachenableAlg0403}.\label{CachenableAlg0408}
\end{algorithmic}
\end{algorithm}
For SFIT, the digital precoder $\mathcal{G}$ and the analog precoding super-matrix $\mathbf{F}_{\mathrm{RF}}$ can be jointly optimized by running Algorithm~\ref{CachenableAlg03} and Algorithm~\ref{CachenableAlg04} in an alternating manner until convergence.
\section*{\sc \uppercase\expandafter{\romannumeral5}. Numerical Results}

In this section, we present numerical results to evaluate the performance of the proposed hybrid precoding algorithms for CeMm-RANs. For simplicity, we assume that all eRRHs have the same number of RF chains $N_{\mathrm{eRF}}$ and the number of antennas $N_{\mathrm{t}}$, respectively. Similarly, all users have the same number of RF chains $N_{\mathrm{uRF}}$ and the number of antennas $N_{\mathrm{r}}$, respectively. All eRRHs have the same maximum transmit power and fronthaul capacity, i.e., $P_{i}=P$ and $C_{i}=C$, $\forall i\in\mathcal{K}_{\mathrm{R}}$. We consider a CeMm-RAN system where the positions of the eRRHs and the users are uniformly distributed within a circular cell of radius $500$ m. The channel model in~\eqref{Cachenable02} is adopted and the average path-loss power is $\rho_{k,i}=\frac{1}{1+\left(d_{k,i}/d_{0}\right)^{\alpha}}$, where $d_{0}$, $d_{k,i}$, and $\alpha$ denote the reference distance, the distance between eRRH $i$ and user $k$ in meters, and the pathloss exponent, respectively. The AoDs/AoAs are assumed to take continuous values and are uniformly distributed in $\left[0, 2\pi\right]$. All users have the same noise variance, i.e., $\sigma_{k}^{2}=\sigma^{2}$, $k\in\mathcal{K}_{\mathrm{U}}$. The eRRHs are equipped with caches of equal size, i.e., $B_{i}=B=\xi SF$, $i\in\mathcal{K}_{\mathrm{R}}$, where $\xi$ denotes the fractional caching capacity. The cache states $c_{f,l,i}$, $f\in\mathcal{F}$, $l\in\mathcal{L}$, $i\in\mathcal{K}_{\mathrm{R}}$, are randomly set subject to $\sum\limits_{f\in \mathcal{F}}\sum\limits_{l\in \mathcal{L}}c_{f,l,i}= \lfloor\xi LF\rfloor $.
If not stated otherwise, the values of the system parameters are as in Table~\ref{SimulatedParametersValues}. For all simulations, the initial digital precoding matrices $\mathbf{G}_{f,l,i}$ and $\mathbf{U}_{f,l,i}$, the analog precoding matrices $\mathbf{F}_{\mathrm{RF},i}$, and the quantization noise covariance matrices $\mathbf{\Omega}_{i}$ are randomly generated and then scaled to satisfy the power and fronthaul capacity constraints.
\begin{table*}[t]
\renewcommand{\captionfont}{\footnotesize}
\renewcommand*\captionlabeldelim{.}
	\setlength{\abovecaptionskip}{0pt}
	\setlength{\belowcaptionskip}{5pt}
	\captionstyle{flushleft}
	\onelinecaptionstrue
	\centering
	\caption{Values of Simulation Parameters.}
	\begin{tabular}{|c|c||c|c||c|c||c|c||c|c||c|c|}
		\hline
		Symbol&\makecell[c]{Value}&Symbol&\makecell[c]{Value}&Symbol&\makecell[c]{Value}&Symbol
		&\makecell[c]{Value}&Symbol&\makecell[c]{Value}&Symbol&\makecell[c]{Value}\\
		\hline
		\hline
		$K_{\mathrm{U}}$& $3$&$K_{\mathrm{R}}$ &$3$&$N_{\mathrm{eRF}}$& $4$&
		$N_{\mathrm{uRF}}$ &$4$&$N_{\mathrm{t}}$& $16$& $N_{\mathrm{r}}$&$16$\\
		\hline
		$F$ & $4$&$\sigma^{2}$&$1$&$L$ & $2$&
		$d_{f,l}$ &$1$&$N_{c}$ &$5$&$\sigma_{k,i}^{2}$ &$1$\\
		\hline
		$d_{0}$ &$50$ m&$\alpha$ &$3$&$d_{\mathrm{a}}$ &$\frac{\lambda_{\mathrm{s}}}{2}$&
		$N_{\mathrm{F}}$ &$2$&$\epsilon$ &$10^{-4}$&$\xi$ &$0.5$\\
		\hline
	\end{tabular}
	\label{SimulatedParametersValues}
\end{table*}

The set $\mathcal{F}_{\mathrm{req}}$ of requested files is randomly generated for each channel realization and all simulation results are obtained by averaging over $1000$ independent channel realizations. For comparison, we also simulate the minimum user rate for fully digital precoding for downlink cache-enabled RANs as a benchmark~\cite{TWCPark2016}. In the legends of the figures, ``Hybrid-HFIT" and ``Hybrid-SFIT" denote the proposed hybrid precoding solutions for HFIT and SFIT, respectively. ``Digital-HFIT" and ``Digital-SFIT" denote the fully digital precoding solutions for HFIT and SFIT, respectively.

\begin{figure}[t]
\renewcommand{\captionfont}{\footnotesize}
\renewcommand*\captionlabeldelim{.}
\centering
\captionstyle{flushleft}
\onelinecaptionstrue
\includegraphics[width=1\columnwidth,keepaspectratio]{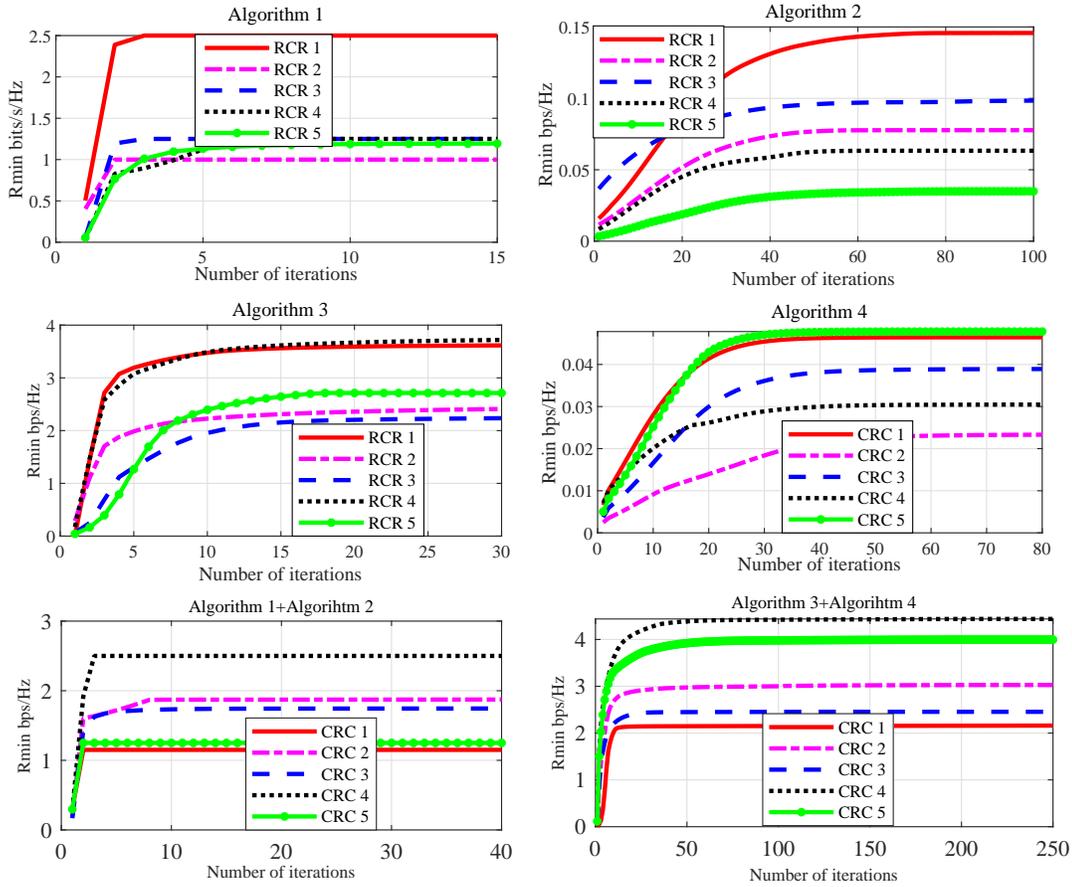}\\
\caption{Convergence trajectories of proposed algorithms for different RCRs.}
\label{AlgorithmsConvergenTrajectory}
\end{figure}

Fig.~\ref{AlgorithmsConvergenTrajectory} illustrates the convergence trajectory of the developed algorithms for different random channel realization (RCR) with $N_{\mathrm{eRF}}=1$, $N_{\mathrm{uRF}}=1$, $L=1$, $S=10$, $P=20$ dB, and $C=2.5$ bit/symbol. Particularly, the numerical results confirm that the adjustment of the radius of the approximation vicinity of the analog precoder in Step~\ref{CachenableAlg0207} of Algorithm~\ref{CachenableAlg02} and Algorithm~\ref{CachenableAlg04} respectively does not change the monotonicity of the generated sequence objective value $R_{\mathrm{min}}$.  Therefore, a non-decreasing sequence of the objective value is obtained with Algorithm~\ref{CachenableAlg01} to Algorithm~\ref{CachenableAlg04} and also by alternating between Algorithms~\ref{CachenableAlg01} and~\ref{CachenableAlg02} (Algorithms~\ref{CachenableAlg03} and~\ref{CachenableAlg04}).

\begin{figure}[t]
\renewcommand{\captionfont}{\footnotesize}
\renewcommand*\captionlabeldelim{.}
\centering
\captionstyle{flushleft}
\onelinecaptionstrue
\includegraphics[width=1\columnwidth,keepaspectratio]{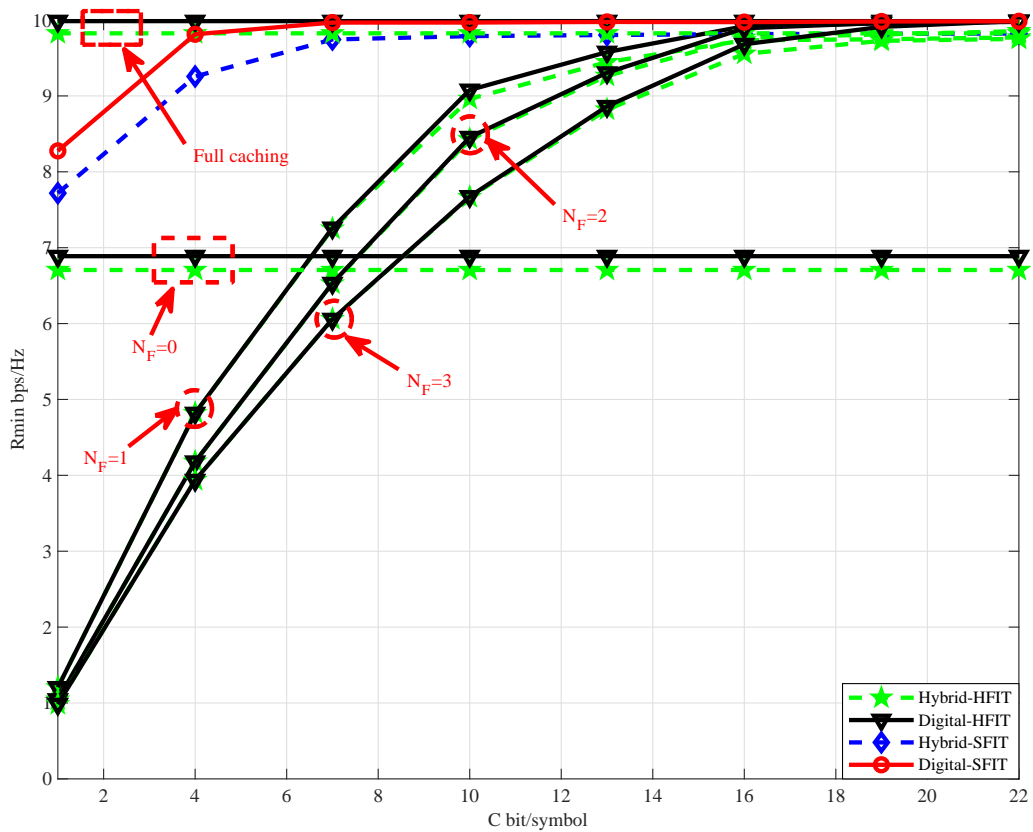}\\
\caption{Minimum user rate versus fronthaul capacity where $N_{\mathrm{eRF}}=2$, $N_{\mathrm{uRF}}=2$, $S=10$, $P=20$ dB, and $K_{\mathrm{U}}=3$.}
\label{CacheFronhaulCapacityComparisonAdd}
\end{figure}

\begin{figure}[t]
\renewcommand{\captionfont}{\footnotesize}
\renewcommand*\captionlabeldelim{.}
\centering
\captionstyle{flushleft}
\onelinecaptionstrue
\includegraphics[width=1\columnwidth,keepaspectratio]{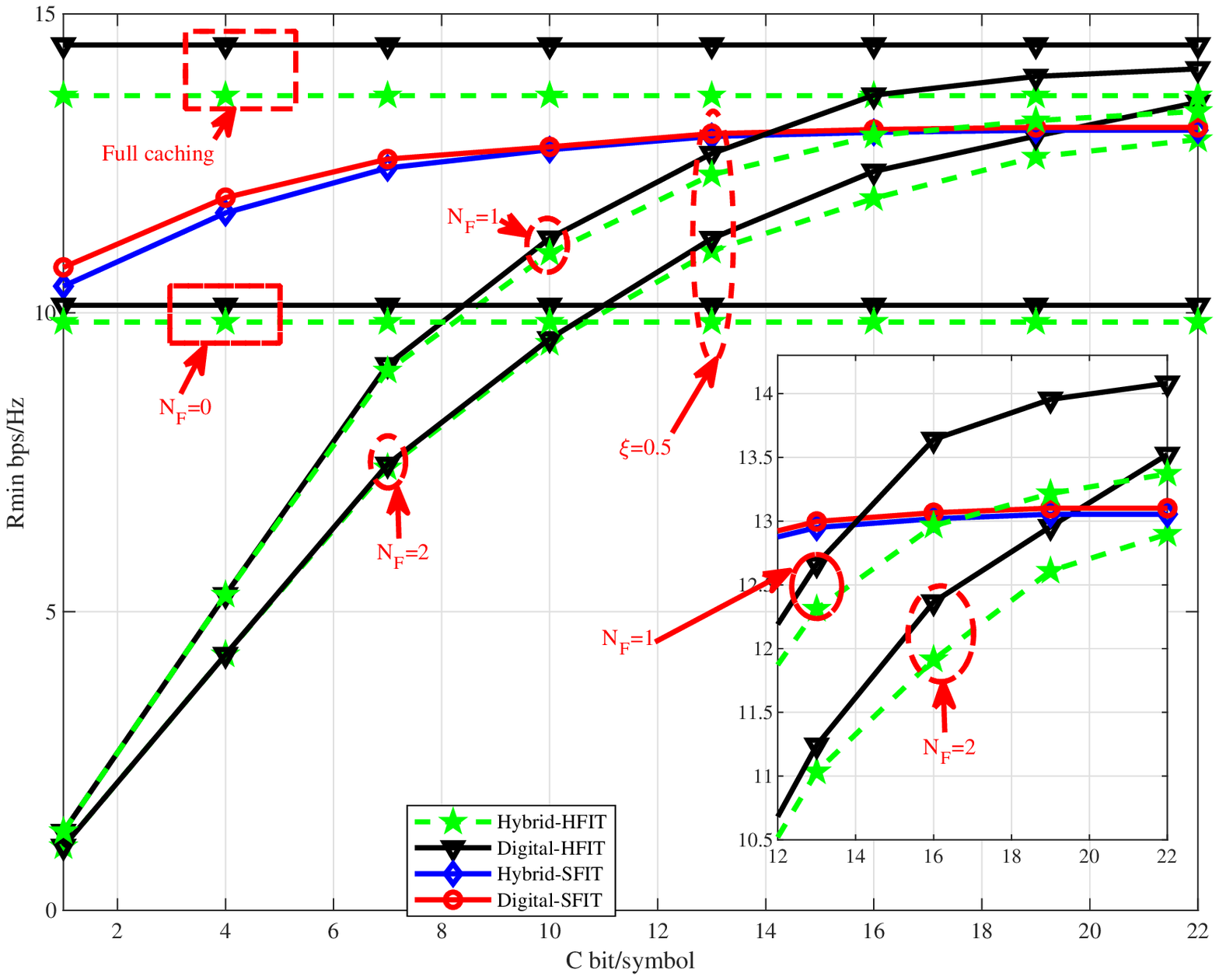}\\
\caption{Minimum user rate versus fronthaul capacity where $N_{\mathrm{eRF}}=4$, $N_{\mathrm{uRF}}=4$, $S=20$, $P=20$ dB, and $K_{\mathrm{U}}=3$.}
\label{CacheFronhaulCapacityComparison}
\end{figure}

\begin{figure}[t]
\renewcommand{\captionfont}{\footnotesize}
\renewcommand*\captionlabeldelim{.}
\centering
\captionstyle{flushleft}
\onelinecaptionstrue
\includegraphics[width=1\columnwidth,keepaspectratio]{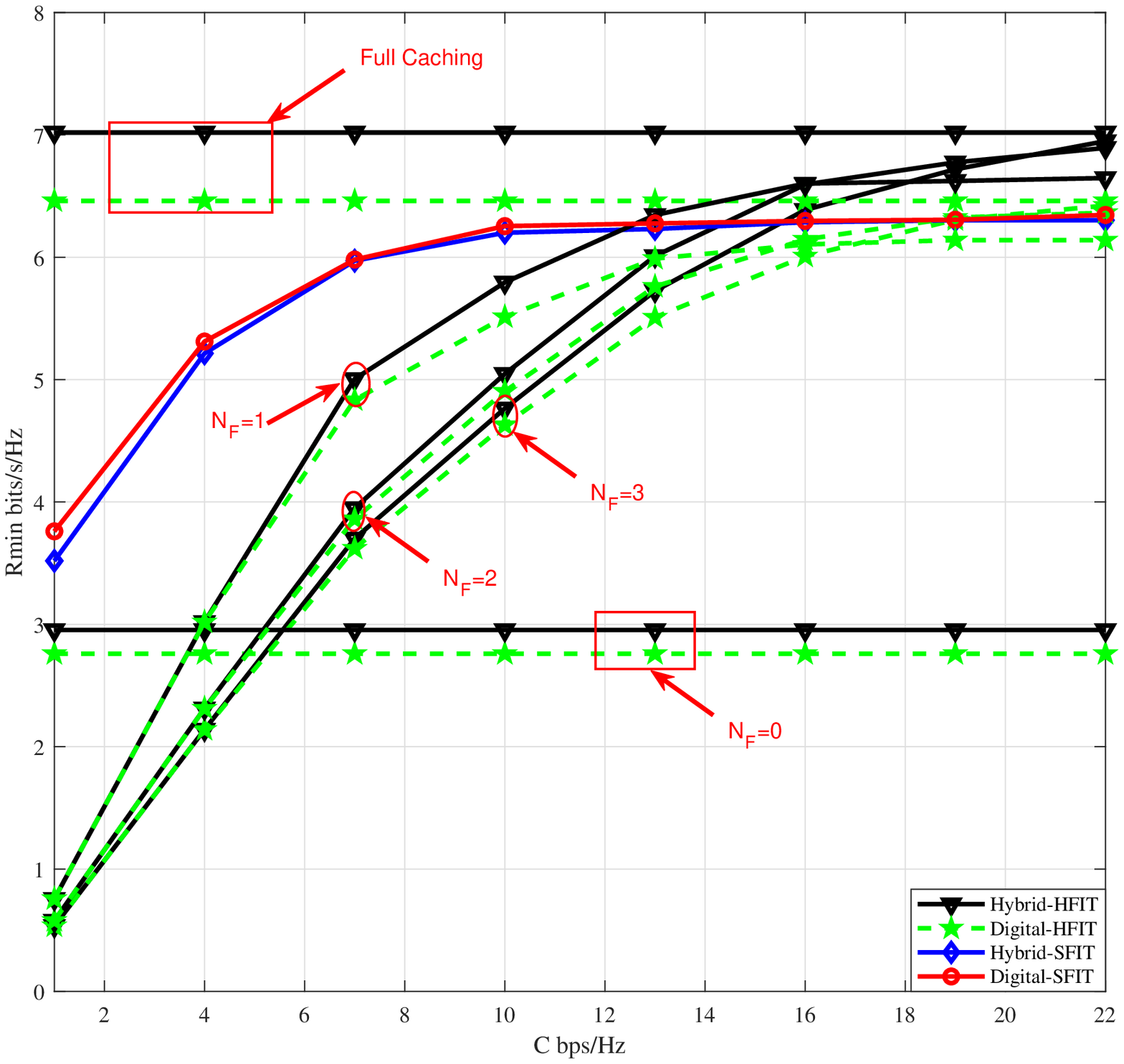}\\
\caption{Minimum user rate versus fronthaul capacity where $N_{\mathrm{eRF}}=4$, $N_{\mathrm{uRF}}=4$, $S=10$, $P=20$ dB, $K_{\mathrm{U}}=8$, and $F=10$.}
\label{LargeNumberofFrontCapacity}
\end{figure}

In Fig.~\ref{CacheFronhaulCapacityComparisonAdd}, Fig.~\ref{CacheFronhaulCapacityComparison}, and Fig.~\ref{LargeNumberofFrontCapacity}, we investigate the average minimum user rate $R_{\mathrm{min}}$ as a function of fronthaul capacity $C$ for downlink CeMm-RANs with different system parameters. By comparing Fig.~\ref{CacheFronhaulCapacityComparisonAdd} and Fig.~\ref{CacheFronhaulCapacityComparison}, one can observe that a larger number of RF chains yields a better system performance. All three figures show that full caching yields the best system performance while the system performance of partial caching is limited by the fronthaul capacity. The minimum user rate $R_{\mathrm{min}}$ first increases with increasing $C$, and saturates when $C$ is sufficiently large. Comparing problems~\eqref{Cachenable12} and~\eqref{Cachenable46}, in addition to constraints~\eqref{Cachenable12b},~\eqref{Cachenable12f}, and~\eqref{Cachenable12g}, for SFIT, the transmit precoders are constrained by the fronthaul capacity $C$ in~\eqref{Cachenable40c} where $\widetilde{g}_{i}\left(\mathcal{U},\mathcal{G},\mathcal{O},\mathcal{Z}\right)$ is a logarithmic function. However, in~\eqref{Cachenable12e}, $\sum\limits_{f\in\mathcal{F_{\mathrm{req}}}}\sum\limits_{l\in\mathcal{L}}d_{f,l,i}R_{f,l}$ is a linear function of $R_{f,l}$. This implies that HFIT is more severely constrained by the fronthaul capacity compared to SFIT. Therefore, SFIT is expected to outperform HFIT in fronthaul capacity limited CeMm-RANs. This observation is confirmed in Fig.~\ref{CacheFronhaulCapacityComparison} for small $C$, e.g. $C<13$ bit/symbol. However, when the fronthaul capacity $C$ is sufficiently large, hybrid precoding with HFIT can outperform hybrid precoding with SFIT. This is because the performance of HFIT is not affected by quantization noise. Also, one can see that for a given fronthaul capacity $C$, the number of coordinated eRRHs $N_\mathrm{F}$ for HFIT should be carefully selected since a larger $N_\mathrm{F}$ requires the transfer of each subfile to more eRRHs on the fronthaul links, which limits the rate of each subfile\footnote{For SFIT, we do not select eRRHs (i.e., parameter $N_{\mathrm{F}}$ does not exist), since in this case the rate on the fronthaul links can be flexibly adjusted via quantization.}. For the cases of $N_\mathrm{F}=0$ and full caching, respectively, the minimum user rate is not constrained by the fronthaul capacity and hence is constant. Besides, one can observe that the average minimum user rate $R_{\mathrm{min}}$ decreases as the number of the users increases by comparing Fig.~\ref{CacheFronhaulCapacityComparison} and Fig.~\ref{LargeNumberofFrontCapacity}. This is because as the number of users increases, the power allocated to each user decreases to maintain the total transmit power constant.  Also, the inter-user interference will increase as the number of users increases.

\begin{figure}[t]
\renewcommand{\captionfont}{\footnotesize}
\renewcommand*\captionlabeldelim{.}
\centering
\captionstyle{flushleft}
\onelinecaptionstrue
\includegraphics[width=1\columnwidth,keepaspectratio]{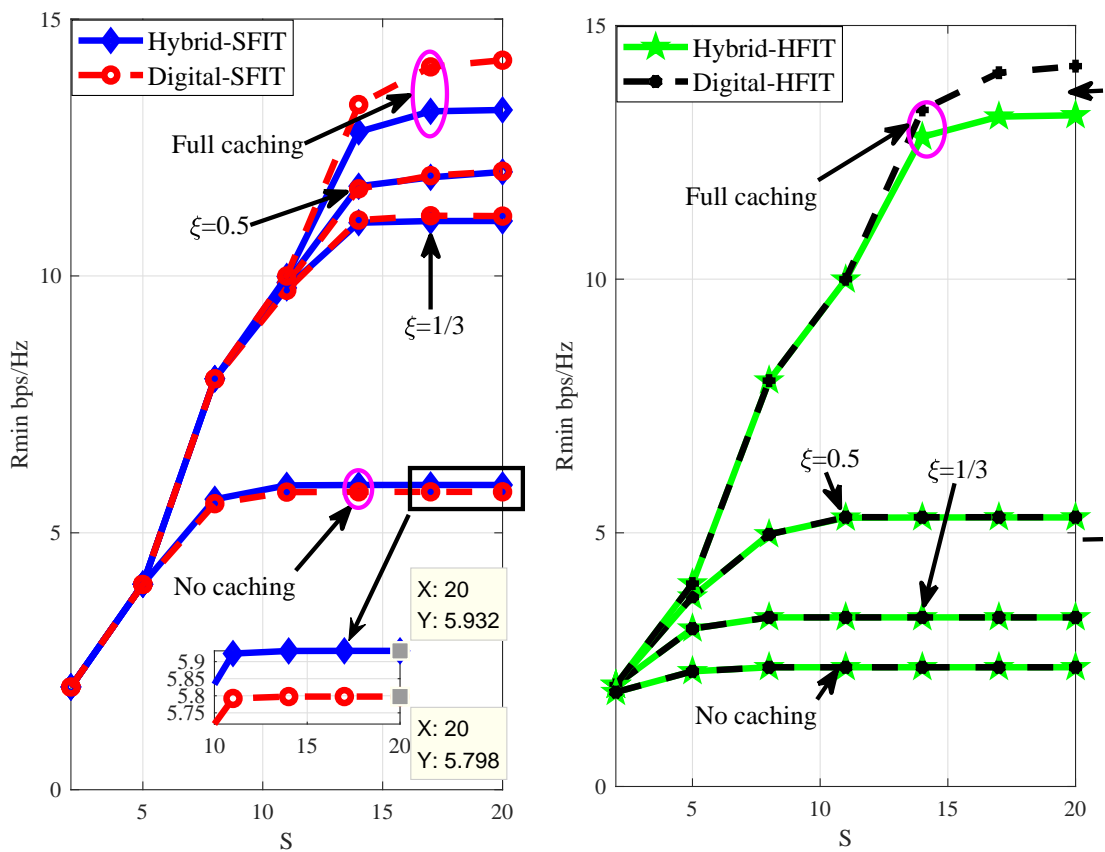}\\
\caption{Minimum user rate $R_{\mathrm{min}}$ versus file size $S$ where $P=20$ dB and $C=5$ bit/symbol.}
\label{CacheFileSizeComparison}
\end{figure}

Fig.~\ref{CacheFileSizeComparison} shows the average minimum user rate $R_{\mathrm{min}}$ versus normalized file size $S$ for downlink CeMm-RANs. As can be observed, for all transmission methods, the minimum user rate $R_{\mathrm{min}}$ increases with increasing $S$ for small normalized files sizes, such as $S=2,\cdots,11$ in our simulations, where the performance is limited by the normalized file size $S$ rather than the fronthaul capacity $C$, the achievable data rate, or the maximum transmit power. In contrast, for large normalized file sizes, the minimum user rate $R_{\mathrm{min}}$ becomes saturated because the performance is limited by the other three factors rather than the normalized file size. When the minimum user rate $R_{\mathrm{min}}$ is limited by the other three factors, SFIT achieves a higher performance than HFIT because the former has more degrees of freedom to adapt to these constraints. We also observe that a larger fractional caching capacity ratio $\xi$ leads to an improved system performance for a given fronthaul transfer strategy. This is because for increasing $\xi$,~\eqref{Cachenable12e} and~\eqref{Cachenable40c}, i.e., the constraints on the fronthaul links, become more relaxed and have less impact on the system performance. Furthermore, different from the results obtained for conventional point-to-point and downlink multiuser mmWave systems~\cite{JSTSPKim2016,TWCEl2014,TWCPark2017,TSPNi2017}, the performance achieved by fully digital precoding is not necessarily better than that of hybrid precoding with SFIT. In particular, for C-RANs with SFIT, i.e., without cache, hybrid precoding outperforms fully digital precoding since the burden on the fronthaul link imposed by transferring uncached requested files increases with the number of antennas, cf.~\eqref{Cachenable35}. In addition, the impact of the quantization noise covariance matrices on the system performance also becomes more serve as the number of antennas increases, cf.~\eqref{Cachenable35} and~\eqref{Cachenable41}.

\begin{figure}[t]
\renewcommand{\captionfont}{\footnotesize}
\renewcommand*\captionlabeldelim{.}
\centering
\captionstyle{flushleft}
\onelinecaptionstrue
\includegraphics[width=1\columnwidth,keepaspectratio]{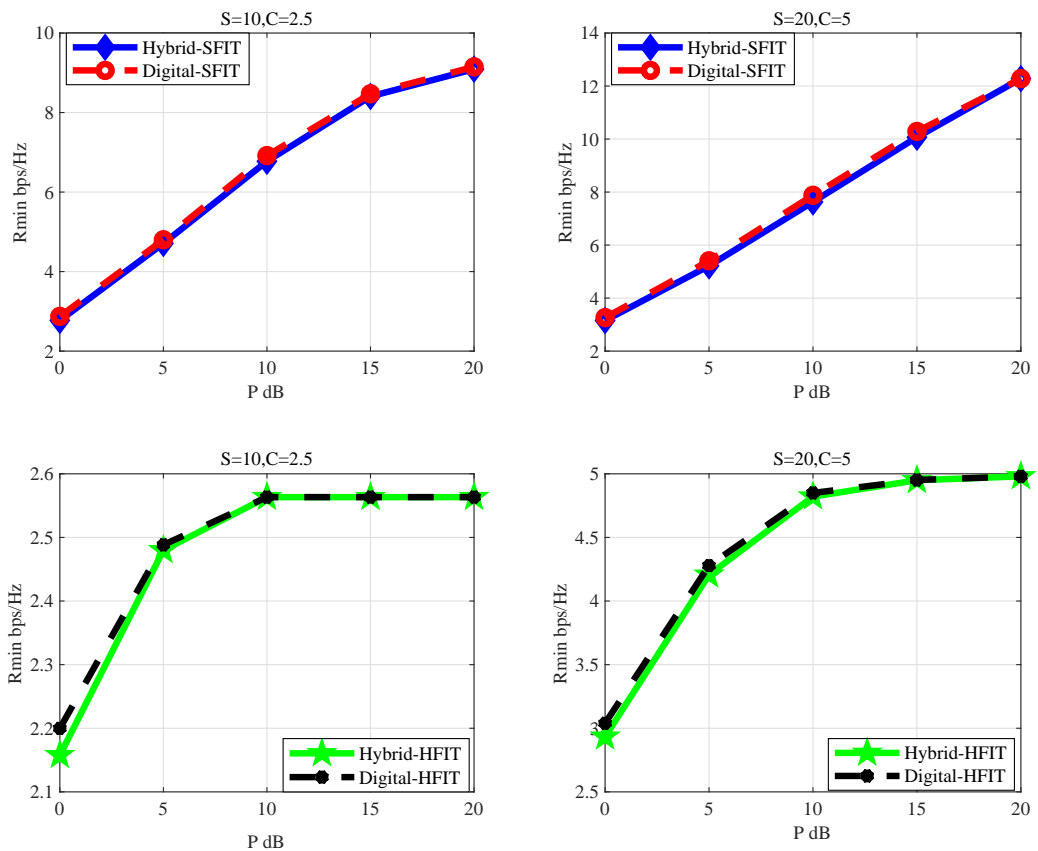}\\
\caption{Minimum user rate $R_{\mathrm{min}}$ versus transmit power constraints.}
\label{CachePowerConstraintsComparison}
\end{figure}

In Fig.~\ref{CachePowerConstraintsComparison}, we show the average minimum user rate $R_{\mathrm{min}}$ versus the maximum transmit power $P$ for CeMm-RANs. One can observe that, for all considered cases, hybrid precoding and fully digital precoding achieve almost the same minimum user rate for CeMm-RANs. Though increasing the transmit power can improve the achievable data rate, the minimum user rate is not only limited by the transmit power constraint, but also by the fronthaul capacity and the caching capacity. Hence, increasing the transmit power does not necessarily improve the system performance in terms of the minimum user rate in a fronthaul capacity or caching capacity limited CeMm-RANs.

\section*{\sc \uppercase\expandafter{\romannumeral6}. Conclusions}
In this paper, we have studied the design of hybrid precoders for CeMm-RANs where each edge node is equipped not only with the functionalities of standard RRHs, but also with local cache and baseband processing capabilities. We considered two basic fronthaul information transfer schemes, i.e., HFIT and SFIT. Specifically, for HFIT, the hard information of uncached files is sent to the eRRHs via the fronthaul links. For SFIT, a quantized version of the precoded signals of the requested uncached files is sent to the eRRHs via fronthaul links. An alternating optimization method was presented to maximize the minimum user rate under constraints on the fronthaul capacity, the eRRH transmit power, and the constant-modulus of the analog precoder. Numerical results were provided to validate the effectiveness of the proposed algorithms and revealed that SFIT achieves a higher performance in CeMm-RANs compared to HFIT in general, except for the case when the fronthaul capacity is large. Furthermore, for medium-to-large file sizes, hybrid precoding with SFIT outperforms fully digital precoding with SFIT for fronthaul capacity limited CeMm-RANs. Interesting topics for future works include the optimization of the clusters of the cooperating eRRHs and the content distribution as well as methods for finding the global optimal solution of the hybrid precoder.

\begin{small}

\end{small}
\end{document}